\documentclass[manuscript,screen]{acmart}
\usepackage[tight,footnotesize]{subfigure}
\usepackage{textcomp}
\usepackage{xcolor}
\usepackage{setspace}
\usepackage{multirow}
\usepackage{epstopdf}
\usepackage{caption}
\usepackage{threeparttable}
\usepackage{extarrows}
\usepackage{setspace}
\usepackage{tabularx}
\usepackage{makecell}
\usepackage{bbding}
\usepackage{pifont}
\usepackage{wasysym}

\newcommand{\tabincell}[2]{\begin{tabular}{@{}#1@{}}#2\end{tabular}}

\AtBeginDocument{%
  \providecommand\BibTeX{{%
    \normalfont B\kern-0.5em{\scshape i\kern-0.25em b}\kern-0.8em\TeX}}}

\acmPrice{15.00}
\acmISBN{978-1-4503-XXXX-X/18/06}

\usepackage[left]{lineno}
\nolinenumbers
\begin{document}
\title{Building Resilient Web 3.0 with Quantum Information Technologies and Blockchain: An Ambilateral View}
\author{Xiaoxu Ren}
\email{xiaoxuren@tju.edu.cn}
\affiliation{%
  \streetaddress{College of Intelligence and Computing}
  \institution{Tianjin University}
  \city{Tianjin}
  \country{China}
  \postcode{300072}
}

\author{Minrui Xu}
\email{minrui001@e.ntu.edu.sg}
\author{Dusit Niyato}
\email{dniyato@ntu.edu.sg}
\affiliation{%
  \streetaddress{School of Computer Science and Engineering}
  \institution{Nanyang Technological University}
  \city{Singapore}
  \country{Singapore}
  \postcode{639798}
}
\author{Jiawen Kang}
\email{kavinkang@gdut.edu.cn}
\affiliation{%
  \streetaddress{School of Automation}
  \institution{Guangdong University of Technology}
  \city{Guangzhou}
  \country{China}
  \postcode{510006}
}
\author{Zehui Xiong}
\email{zehui_xiong@sutd.edu.sg}
\affiliation{%
  \streetaddress{Pillar of Information Systems Technology and Design}
  \institution{Singapore University of Technology and Design}
  \city{Singapore}
  \country{Singapore}
  \postcode{487372}
}
\author{Chao Qiu}
\email{chao.qiu@tju.edu.cn}
\author{Xiaofei Wang}
\email{xiaofeiwang@tju.edu.cn}
\affiliation{%
  \streetaddress{College of Intelligence and Computing}
  \institution{Tianjin University}
  \city{Tianjin}
  \country{China}
  \postcode{300072}
%\postcode{* Corresponding author}
}


\renewcommand{\shortauthors}{Ren and Xu et al.}

\begin{abstract}
% Web 3.0 seeks to establish decentralized ecosystems utilizing blockchain technologies to drive the digital transformation of physical commerce and governance. Blockchain-based consensus algorithms and smart contracts, grounded in cryptographic technologies, enable digital identity, digital asset management, decentralized autonomous organizations, and decentralized finance. These elements create secure and transparent digital economy services in Web 3.0, fostering the integration of digital and physical economies. Concurrently, the rapid development of quantum devices has led to the emergence of quantum cloud computing and quantum internet alongside Web 3.0. Quantum computing initially poses challenges to traditional cryptographic systems that protect data security, while also reshaping modern cryptography with its inherent advantages. In this survey, we offer a comprehensive overview of the blockchain-based Web 3.0 landscape, as well as its quantum and post-quantum enhancements, from an ambilateral perspective. We examine post-quantum migration methods and anti-quantum signatures, which provide potential avenues for achieving unforgeable security under quantum attack for blockchain's internal technologies. Simultaneously, we explore quantum/post-quantum encryption and verification algorithms that improve blockchain's external performance, paving the way for a decentralized, valuable, and secure system. Finally, we discuss future directions in developing a provably secure decentralized digital ecosystem.
Web 3.0 pursues the establishment of decentralized ecosystems based on blockchain technologies to drive the digital transformation of physical commerce and governance. Through consensus algorithms and smart contracts in blockchain, which are based on cryptography technologies, digital identity, digital asset management, decentralized autonomous organization, and decentralized finance are realized for secure and transparent digital economy services in Web 3.0 for promoting the integration of digital and physical economies. With the rapid realization of quantum devices, Web 3.0 is being developed in parallel with the deployment of quantum cloud computing and quantum Internet. In this regard, quantum computing first disrupts the original cryptographic systems that protect data security while reshaping modern cryptography with the advantages of quantum computing and communication. Therefore, this survey provides a comprehensive overview of blockchain-based Web 3.0 and its quantum and post-quantum enhancement from the ambilateral perspective. On the one hand, some post-quantum migration methods, and anti-quantum signatures offer potential ways to achieve unforgeable security under quantum attack for the internal technologies of blockchain. On the other hand, some quantum/post-quantum encryption and verification algorithms improve the external performance of the blockchain, enabling a decentralized, valuable, secure blockchain system. Finally, we discuss the future directions toward developing a provable secure decentralized digital ecosystem. 
\end{abstract}




\maketitle

\section{Introduction}

\subsection{Background }

The decentralized digital economies in Web 3.0, including creator economies and decentralized autonomous organizations (DAOs), can improve the user experience and take user privacy and security to the next level~\cite{chen2022digital, Quantumxu}. Without trusted authorities, users own their unique digital identities built on blockchain to access personal data and decentralized applications in Web 3.0. In addition to read-and-write operations, Web 3.0 can provide proof-of-ownership for digital assets of participants based on blockchain technologies. In Web 3.0, users' data is stored in or linked to the blockchain and stored in distributed data storage systems such as the InterPlanetary File System (IPFS)~\cite{daniel2022ipfs}. Meanwhile, activities and regulations in Web 3.0 are executed via immutable and transparent smart contracts and consensus algorithms proposed by DAOs.

Web 3.0's decentralized nature empowers users with greater control over their data, which reduces reliance on the management of centralized authorities~\cite{chen2022digital}. The role of blockchain is to enhance the security and transparency of Web 3.0 by providing a trustworthy way for users to store data and conduct transactions without the need for centralized authorities. However, it is essential to note that blockchain technology is susceptible to quantum computing, which can break one-way mathematical functions such as hash functions and public-key encryption commonly used in classical computing. Due to the superposition and entanglement of quantum bits (qubits), quantum computing has enough processing power to break existing digital signatures and hash functions within secrecy periods. For example, a quantum computer with 2$\times$10$^7$ qubits can crack RSA-2048 in just eight hours~\cite{gidney2021factor}. Hence, it is possible for a malicious user utilizing quantum computers to easily steal private keys from a blockchain wallet, ultimately resulting in the loss of all funds in that wallet.

While Web 3.0 affords users the ability to peruse, produce, and possess their user-generated content (UGC), traditional blockchain technology is on the verge of obsolescence once the potency of a quantum computer meets the prerequisites to operate Shor's algorithm and Grover's algorithm~\cite{fedorov2018quantum}. The proliferation of quantum information technologies has opened a new avenue for blockchain technology in Web 3.0. A quantum blockchain, which is founded on quantum information technologies, can formulate more secure and resilient distributed ledgers that are impervious to quantum attacks. For instance, quantum key distribution (QKD) protocols encode the key employed for symmetric encryption into a quantum state for transmission over a quantum channel. With one-time-pad technology, QKD-secured communication can realize information-theoretically secure data transmission. Furthermore, quantum blockchain can utilize the quantum hash function and one-way quantum computing functions to fabricate quantum voting~\cite{pirandola2020advances} and quantum signature~\cite{wang2022quantum} algorithms. Additionally, post-quantum blockchain, which is based on post-quantum cryptographic algorithms that are immune to quantum attacks, can also be deemed as a solution to secure Web 3.0 in the post-quantum era. Consequently, classical blockchain is revolutionized into the quantum blockchain with the added benefits of prolonged data encryption and over two decades of usage.

Due to the substantial current interest and complementary advantages of quantum information technologies and blockchain, we aim to examine how state-of-the-art quantum information technologies are driving blockchain to open up new horizons for the development of Web 3.0 from the ambilateral perspective. Specifically, we explore the benefits of quantum blockchain in terms of internal technologies, i.e., consensus protocols, incentive mechanisms, smart contracts, and cryptographs. In addition, we also investigate how quantum technologies can enhance the performance of blockchain in terms of external aspects such as decentralization, profitability, and privacy of storage.

% 第一段: From Blockchain to Web 3.0

% 第二段: The development of quantum computing threatens the 经典密码学 in blockchain

% 第三段: To 维护 Web 3.0 中的去中心化, 安全 和 透明, 量子密码学和后量子密码学

% 第四段: 

% 第五段: 

% Why this survey is important for my research

% After I read this paper, what research we can do

%\textcolor{blue}{Recently, immense volume of data are generated approximately by the mega-scale terminal devices instead of traditional cloud datacenters, such as, mobile devices, vehicles and other smart devices.}

%\textcolor{blue}{Quantum, as the terminal intelligence, is a burgeoning paradigm integrated network, computing, storage and AI, providing EI services and satisfying the key requirements of internet era in agile connection, real-time business, data optimization, application intelligence, security and privacy protection, etc. Deploying intelligence on edge devices could make intelligence be closer to users and provide AI services for users faster and better. }

%\textcolor{blue}{However, }

%\textcolor{blue}{On the other hand, blockchain (BC), as the underlying technology of crypto-currencies, has been a relatively recent technological trend. }

%\textcolor{blue}{Indeed, the quantum blockchain synthesizes decentralized data storage, computing and BC-based decentralized security. }

\begin{figure*}[!t]

\centering

\includegraphics[width=14cm]{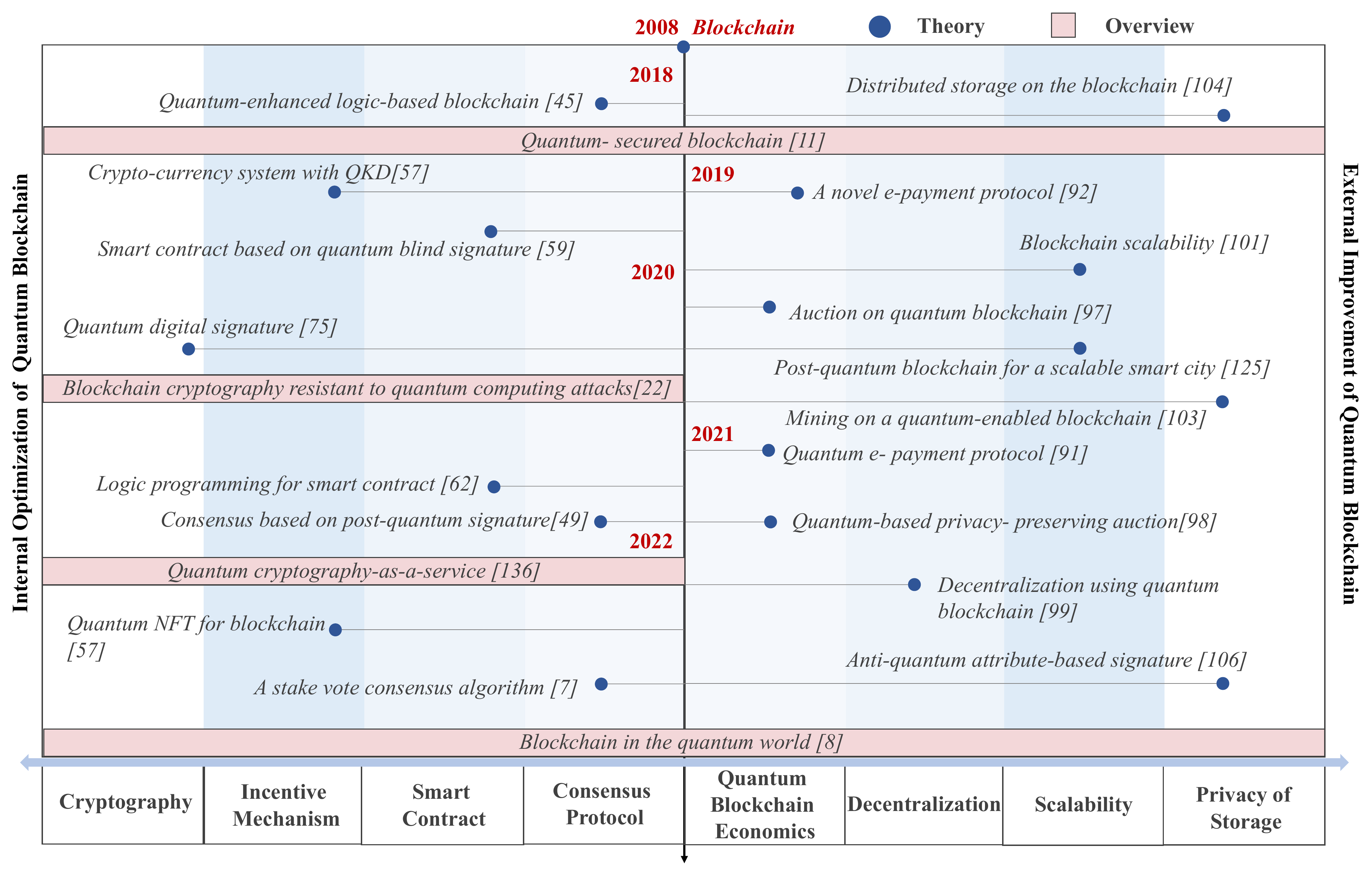}

\caption{The research activities of Quantum-Driven Blockchain in Web 3.0}

\label{fig:activities}

\end{figure*}
\subsection{Related Books and Surveys}

Recently, there have been many efforts that attempt to converge quantum information technologies and blockchain into the quantum and post-quantum blockchain. Meanwhile, the growth of quantum information theory and computation has resulted in a rise in the number of research ongoing in the quantum blockchain. 

Some works on frameworks, opportunities, and challenges of quantum blockchain are \cite{Faridi_2022,krishnaswamy2020quantum,li2019quantum,Kiktenko_2018,quantum1010002,Dey21challen,Allende21}.
Faridi \emph{et. al} explore the possibilities and potential benefits of quantum computing in blockchain systems \cite{Faridi_2022,krishnaswamy2020quantum}. The advantages of quantum blockchain and development prospects are summarized in \cite{li2019quantum} compared with the classical blockchain while providing the detailed method of applying quantum technology to the blockchain. Kiktenko \emph{et. al} give an overview of the current state of research on quantum blockchain \cite{Kiktenko_2018}, covering topics such as quantum-resistant blockchain protocols and cryptography algorithms. Vlad \emph{et. al} explore the potential of using entanglement in time to create a quantum blockchain and present the challenges of implementing a quantum blockchain \cite{quantum1010002}. Further, Dey \emph{et. al} provide an overview of the challenges and opportunities associated with developing a quantum blockchain \cite{Dey21challen}, including issues related to quantum security, scalability, and interoperability. Allende \emph{et. al} propose some tutorials on the implementation of quantum blockchain \cite{Allende21}, where quantum entropy was provided via the IronBridge Platform from CQC and LACChain Besu is used as the blockchain network. However, the above works only focus on either internal technical optimization (i.e., consensus protocols, incentive mechanisms, smart contracts, and cryptography) or external performance improvement (i.e., economy, decentralization, scalability, and privacy of storage). There is no research that explores the benefits of quantum blockchain from internal optimization and external performance. Therefore, this survey provides a comprehensive account of quantum blockchain from an ambilateral View to build resilient Web 3.0 infrastructure. We summarize the research activities of quantum-driven blockchain in Web 3.0, which is shown in Fig. \ref{fig:activities}. 

Unless quantum technology is integrated into blockchain technology, the components of classical blockchain are exposed to quantum attacks \cite{fedorov2018quantum}, resulting in security risks to the internal technologies and external performance of the blockchain. Currently, we have already seen the development of information-theoretically secure quantum protocols, such as quantum homomorphic encryption protocols \cite{liu2022efficient,ouyang2018quantum}, quantum image encryption \cite{li2020quantum,li2019attacker}, and quantum digital signature \cite{yin2016practical,zeng2002arbitrated}, while several attempts have been made to incorporate post-quantum cryptography \cite{grote2019review,fernandez2020towards}. These protocols 
bring benefits to the internal technologies and external performance of blockchain, achieving privacy, security, effectiveness, and profitability under quantum attacks.

Nevertheless, the above studies focus on either internal technical optimization or external performance improvement of quantum technologies for blockchain in Web 3.0. There is no research that explores the benefits of quantum technology for blockchain in both aspects. Therefore, this survey provides a comprehensive account of blockchain combined with quantum information technology from both internal and external aspects in a Web 3.0 scenario, starting from an overview, motivation, and comprehensive framework to research challenges and future directions.

\begin{figure*}[!t]

\centering

\includegraphics[width=14cm]{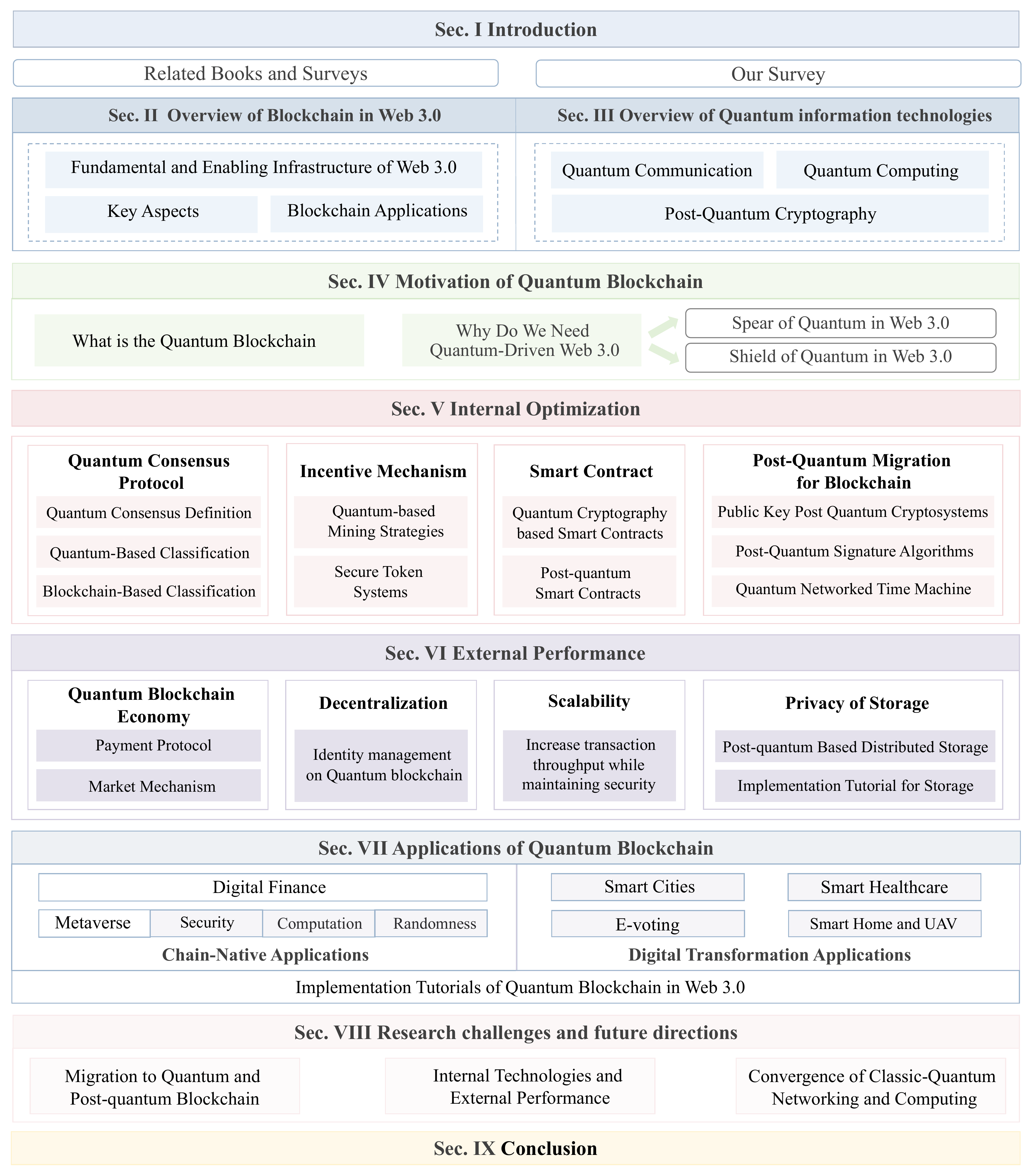}

\caption{The taxonomy graph of Quantum-Driven Blockchain in Web 3.0}

\label{fig:taxonomy}

\end{figure*}

%\textcolor{blue}{This survey is motivated due to the current eruption of interest around Quantum technology and BC. Although these two technologies have been studied extensively in previous works, there are few works to survey the conjunction between them. Therefore, we examine whether the integration of EI and BC will make a powerful network with combined functionalities of both cutting-edge technologies. }

\subsection{Our Survey}

Advances in quantum computers undermine the security of classical blockchains, there is an urgent need to solve the limitations of blockchain and continue to develop Web 3.0.  This paper aims to integrate quantum information technologies into blockchain and investigate the most recent developments in Web 3.0 thoroughly. The contributions of this survey can be summarized as follows:

\begin{itemize}
	\item First, we present the basic principles of blockchain and quantum information technology, while giving motivations for using the complementary properties of quantum technology to support blockchain from the ambilateral perspective, i.e., internal technologies optimization and external performance improvement.
	\item Second, we optimize the internal technologies of quantum blockchain from four aspects, including consensus protocols, incentive mechanisms, smart contracts, and cryptography. 
	\item Third, we improve the external performance to better support quantum blockchain, including economy, decentralization, scalability, and privacy of storage.
	\item Fourth, we investigate potential quantum blockchain applications and give tutorials for implementing quantum blockchain in Web 3.0.
	\item Finally, we explore several key challenges and open research directions.
\end{itemize}

A taxonomy graph of this paper is presented as Fig. \ref{fig:taxonomy}. Specifically, we first give the background on blockchain and quantum information technologies in Section \ref{Overview_Blockchain} and Section \ref{Overview_Quantum}. Then, the motivation for integrating them is explained in Section \ref{Motivation}. Next, we describe in detail the integration techniques from internal technologies optimization (Section \ref{BC}) and external performance improvement (Section \ref{Edge}). In Section \ref{Applications}, we investigate some applications and tutorials of quantum blockchain. Section \ref{Challenge} discusses research challenges and future directions. Lastly, the conclusion is drawn in Section \ref{Conclusions}.

%\textcolor{blue}{To the best our knowledge,  the integration of EI and BC will gage some problems being faced by edge devices as BC provides advanced features including persistency, traceability, anonymity, integrity, security, etc., while storing information in a public decentralized but privacy-preserving manner. }

%\textcolor{blue}{A taxonomy graph of this paper  is presented as Fig. \ref{fig:architecture}. Then, it introduces the background knowledge of edge computing, distributed learning and BC. The motivations of the BC-enabled EI with distributed learning is explained in Section III. Meanwhile, the functions of this integration are presented in Section IV. Moreover, we review the recent developments of BC-enabled EI in Section V with extensive discussion of benefits of BC-enabled EI models in a wide range of advanced applications. Final, research challenges and future directions are outlined in Section VI.}

\section{Overview of Blockchain in Web 3.0}
\label{Overview_Blockchain}
\begin{figure*}[!h]

\centering

\includegraphics[width=15cm]{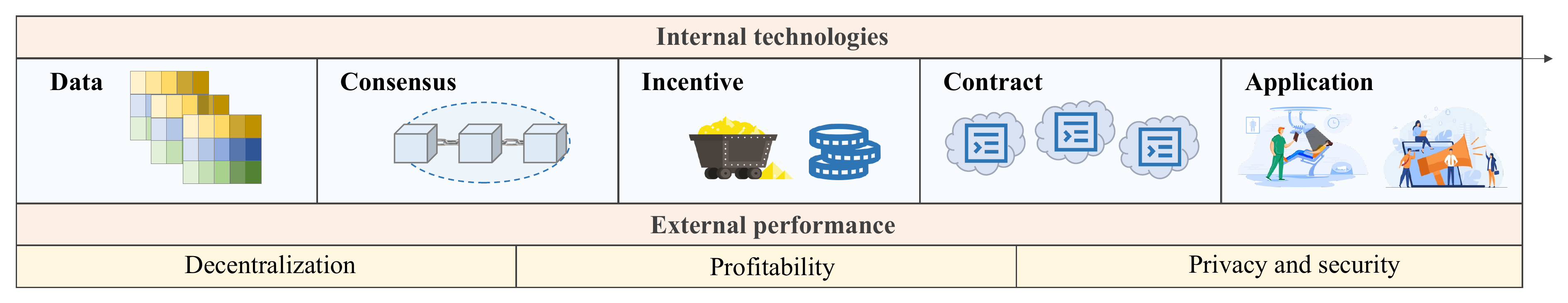}

\caption{The background of blockchain in Web 3.0}

\label{fig:background}

\end{figure*}

In this section, we present the background of blockchain in Web 3.0, including the definition, internal technologies, and external performance, which is shown in Fig. \ref{fig:background}.
\subsection{Fundamental and Enabling Infrastructure of Web 3.0}
As the cornerstone of the digital economy, Web 3.0 can drive digital transformation in a decentralized, secure, and transparent way. The read/write/own Web 3.0 is evolved from the read Web 1.0 and the read/write Web 2.0, which allows content creators and users to own the rights to use and manage their digital assets~\cite{chen2022digital}. In Web 1.0, users could only read online content published by authoritative publishers who own most of the data and value on the Internet. The data in Web 1.0 is stored on the local servers of Internet owners. Transited to Web 2.0, users can access information from online applications and platforms as well as create and share their content with other users. However, the value of data is stored on cloud servers and owned by companies, and users still cannot own the content they read and create. Finally, in Web 3.0, users access the Internet with their digital identities and wallets, which are leveraged to manage the data they create in online services. Therefore, Web 3.0 is also conceived as the Internet of Value~\cite{alabdulwahhab2018web}, which allows users to create, manage, and trade their user-generated content in the form of digital assets. 

The blockchain is developed based on consensus algorithms and smart contracts that can provide decentralization and interoperability for Web 3.0 applications and services~\cite{zheng2018blockchain}. Each user in Web 3.0 owns a digital wallet that consists of its digital identities. Through digital wallets, users can interact with distributed ledgers on Web 3.0 to manage their behavioral data assets, digital assets, and transaction histories, which are recorded in the on-chain database anonymously. Distributed data storage enables Web 3.0 applications to store user interaction and authentication data that previously had to be stored on distributed data servers. During the utilization, users gain access to the data through their digital identities. The benefits of storing data via distributed data storage systems include low storage costs, high security, and low storage costs~\cite{daniel2022ipfs}. Distributed storage uses smart contracts to distribute data across multiple network nodes and fully utilize their storage resources.

%Web 3.0中的用户拥有标志他们数字身份的数字钱包， 并通过钱包与区块链中的分布式账本中进行交互与记录。 Specifically, 用户的资产，交易记录 行为信息和用户数据都可以被完全匿名的记录在分布式数据存储中。

% Computing-power networks:

% Blockchain technologies:

% The seamless connectivity facilitated by ubiquitous networking infrastructure offers significant advantages by enabling the convergence of computing and networking resources located in geographically diverse areas. This computing and networking convergence results in the formation of a decentralized computing and storage infrastructure, which provides the necessary resources to support various Web 3.0 services. By merging computing resources from end devices, edge servers, and cloud computing platforms (including classical and quantum cloud computing), CPNs can facilitate a wide range of use cases, such as record-keeping, proof of ownership, scheduling, and trading of decentralized digital finance and marketplaces. The use of distributed storage technology in decentralized applications enables the storage of massive amounts of data and digital assets across distributed storage nodes, along with recording the data retrieval links used for data identification. Additionally, blockchain-based dApps can serve as distributed ledgers for digital identities and assets, providing unified access to all Web 3.0 participants. Overall, the seamless integration of different technologies provides a robust and secure infrastructure to support various use cases for Web 3.0 applications.

\subsection{Key Aspects of Blockchain in Web 3.0}

As mentioned above, blockchain-based digital ecosystems with distributed data storage and computing-power networks are the foundation for the digital society of Web 3.0. Notably, Web 3.0 is considered to be a collection of blockchain-based protocols focused on changing the ecology of the Internet. This survey describes the key aspects of blockchain in Web 3.0 in terms of internal technologies and external performance.

\subsubsection{Internal Technologies} It refers to the support of blockchain for Web 3.0 in terms of internal composition.

%\textcolor{blue}{Generally, the BC architecture is decoupled into key internal technologies \cite{7795984,7950404}, as shown in the following. }

\begin{itemize}

\item \emph{Consensus Protocols:} Consensus protocols are an integral part of the decentralized Web 3.0, which are leveraged to establish consensus among multiple interest parties without trusted third-party authorities. This ensures the security and reliability of the network. Several consensus protocols of blockchain in Web 3.0 include proof of work (PoW) \cite{gervais2016security}, proof of stake (PoS) \cite{king2012ppcoin}, proof of authority (PoA) \cite{de2018pbft}, etc. Different consensus protocols have different properties and are applicable to different Web 3.0 applications and services. 

\item \emph{Incentive Mechanism:} As the Internet of Value, Web 3.0 is built on a plethora of favorable incentives for encouraging participants. There are multiple types of incentives, e.g., miner rewards, transaction gas, and minting fees, distributed based on blockchain platforms in Web 3.0. In addition, incentives in blockchain technologies allow participants for the creation and management of digital assets as NFTs \cite{voshmgir2020token}. This function of the blockchain provides the fundamental of Web 3.0 \cite{murray2023promise}, which involves the tokenization of various content, including digital identities, intellectual properties, and real-world assets.

\item \emph{Smart Contract:} As the foundation of Web 3.0 and the new digital environment, smart contracts are codes of digital auto-executing applications. These contracts support a range of functions, from executing financial transactions to identifying users to running decentralized applications. Smart contracts govern the rules of Web 3.0 agreements \cite{cao2022decentralized}. They allow users to interact with dAPPs in Web 3.0 by the blockchain. 
Gaming applications, in particular, are booming with the advent of Web 3.0 and Metaverse. Smart contracts in Web 3.0 combine tokens or crypto rewards with gameplay by the GameFi and Play to Earn. 
Thus, smart contracts are opening a new era of combined entertainment and economy in Web 3.0.

\item \emph{Cryptography:}   
Cryptography is a method of storing and transmitting data in a specific form that prevents third parties from accessing and getting information from private messages during peer-to-peer communication. Generally, this aspect of blockchain is expected to be critical to Web 3.0 to protect user data and privacy \cite{wang2022exploring}.

\end{itemize}

\subsubsection{External Performance} It refers to the support of blockchain for Web 3.0 in terms of external properties.

\begin{itemize}
\item \emph{Decentralization:} As mentioned above, one of the core issues of Web 2.0 is the centralized management of authority and data. Blockchain decentralizes Web 2.0 into Web 3.0 by facilitating a broader distribution of data and authority. Web 3.0 adopts the blockchain-driven public distributed ledger for greater transparency and decentralization, creating a more open, transparent, and secure Internet.

\item \emph{Profitability:} Web 3.0  disrupts the DAPP industry through tokenizing assets. This involves the creation of digital tokens, which can be bought, sold, and traded on blockchain-based platforms, making it easier for investors to diversify their portfolios and gain more profits. 

\item \emph{Privacy and Security:} Blockchain plays a crucial role in the privacy and security of data storage and identity management. From Web 1.0 to Web 2.0, significant progress are made in decentralizing data, empowering users to create and share data, whereas users do not own and control their data. Further, Web 3.0 provides users with an open, trust, and permissionless Internet to ensure data ownership and privacy.

\end{itemize}

\emph{\textbf{Lessons learned:}} According to the description above, it is quite clear that blockchain can serve as a key driver of the next generation of the Internet. As an emerging technology, blockchain brings many advantages while it also needs to explore in tackling its disadvantages. Transactions on the blockchain are public and transparent, making it difficult to protect personal privacy. Each blockchain nodes save whole data, causing network congestion and slow transaction speed. Moreover, reaching a consensus consumes a lot of resources consumption. Especially with the arrival of quantum technology, blockchain would face more restrictions. In the next section, we focus on the background of quantum information technologies.

\section{Overview of Quantum Information Technologies}
\label{Overview_Quantum}

\begin{figure*}[!h]
\centering
\includegraphics[width=15cm]{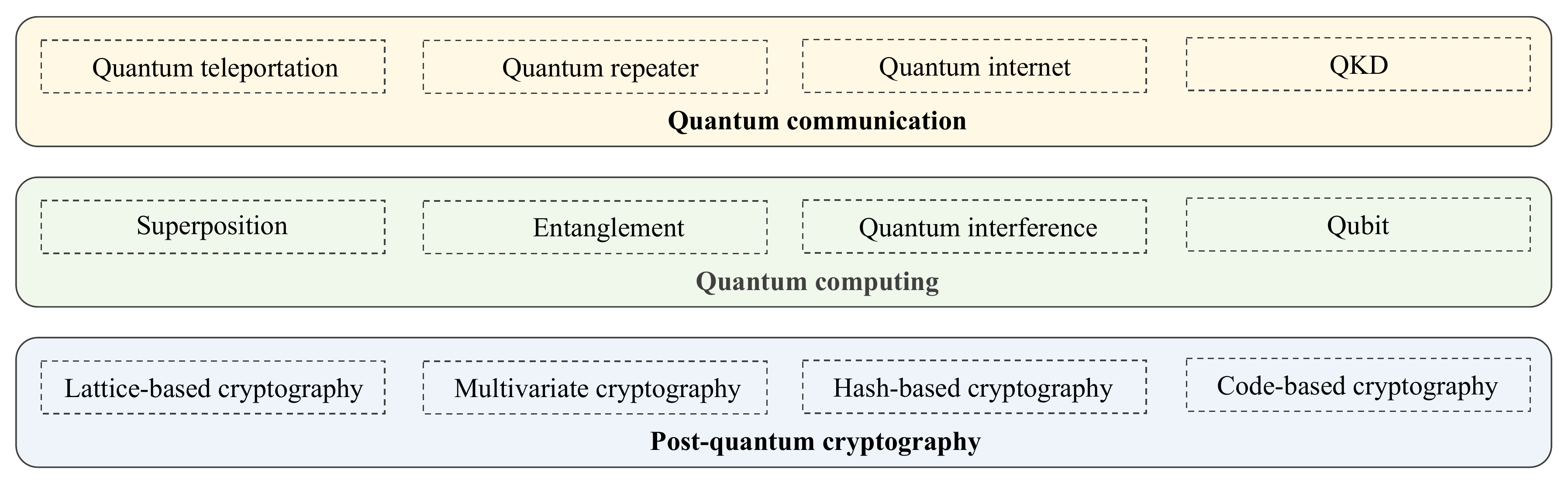}
\caption{The background of quantum information technologies in Web 3.0}
\label{fig:quntum}
\end{figure*}

In this section, we give the background of quantum information technologies, including quantum communication, quantum computing, and post-quantum cryptography, which is shown in Fig. \ref{fig:quntum}.
\subsection{Quantum Communication} 
In the Internet of Everything era, huge amounts of sensitive data rely on communication networks. These data are usually encrypted and then sent along with a ``key" over fiber-optic cables. Specially, these data and keys are sent as classical bits, which can be read and copied by hackers without leaving a trace, thus causing security problems. In the following, we give some key terms to fully understand quantum communication.

\begin{itemize}
\item \emph{Quantum:} 
Quantum in ``quantum computing" means the quantum mechanics used by the system to calculate the output. Generally, a quantum refers to the smallest possible discrete unit of any physical property, e.g., electrons, neutrinos, and photons.

\item \emph{Qubit:} 
Similar to the bits in classical computing, qubits are the basic unit of information in quantum computing. The difference between them is that a classical bit is binary and it can only take the value of 1 or 0, while a qbit can maintain a superposition of all possible states.
\end{itemize}

Quantum communication uses the laws of quantum physics to protect private data \cite{Gisin07NP}. These laws allow particles, i.e., photons that transmit data, to take on a superposition state. That is the result of measurement of such states is not definite $|0\rangle$ or definite $|1\rangle$, where $|\cdot\rangle$ is called a “ket-vector” in Dirac notation. It also means particles can take on multiple combinations of values of 1 and 0 simultaneously. These particles are called quantum bits, or qbits. The super-fragile quantum state of qubits causes hackers to be unable to tamper with them without leaving signs of activity. Quantum communication involves the generation and use of quantum states for communication protocols. 

\subsubsection{Quantum Teleportation} 
Quantum teleportation is a process by which a qubit can be transmitted from a sender to a receiver, without that qubit actually being transmitted through space \cite{pirandola2015advances}.
Unlike the portrayed teleportation in science fiction, which is described as a means of transferring physical objects from one location to another, quantum teleportation only transmits quantum information \cite{zeilinger2000quantum}. Moreover, quantum teleportation exploits the property of quantum mechanics, i.e., quantum entanglement to transfer the quantum state of a particle onto a different particle \cite{furusawa1998unconditional}. 

\subsubsection{Quantum Key Distribution} 

Quantum communication is primarily concerned with secure communication by developing secure quantum channels. One of the most interesting topics in this field is QKD, which uses quantum teleportation to generate key pairs that allow encryption and decryption of messages \cite{RevModPhy8109}.

Currently, there emerges a number of  approaches and protocols for implementing QKD. QKD prevents hacks and provides excellent privacy protection while addressing the key distribution problem based on quantum cryptography \cite{Diamanti16}. We use one known BB84 as an example to introduce the QKD process. Assume that Alice would like to send data securely to Bob.

\begin{itemize}
\item \emph{Generation and transmission of encryption key:} Alice creates an encryption key in the form of qubits. Then, these qubits would be sent to Bob by the fiber-optic cable.
\item \emph{State measurements of qbits:}
Bob then uses a random base to measure the qbits received with the basis, after which Bob acknowledges to Alice the quantum bits he received through the public channel.
\item \emph{Key sifting:} Alice and Bob can determine that they hold the same key by comparing the state measurements of a fraction of the qbits.
\item \emph{Key distillation:} It is run by Alice and Bob, involving calculating whether the error rate is high enough to indicate that the hacker is trying to intercept the key.
\item \emph{Key update:}  If so, the suspect key might be threw away while the new keys are generated until Alice and Bob are confident that they share a secure key. Then, Alice can encrypt the data with her key and send it as classical bits to Bob, who uses his key to decode the message.
\end{itemize}

\subsection{Quantum Computing}

Classical computers today encode information into bits using a binary stream of electrical pulses (1 and 0), which restricts their processing ability. Compared to classical computers, quantum computers use quantum qubits, which can take a value of 1 or 0, or a complex combination of both 1 and 0, to perform calculations that ordinary bits can't. Quantum computers are capable of measuring and observing quantum systems at the molecular level, as well as solving conditional probabilities of events. They serve to accelerate the development of artificial intelligence and web 3.0. 

Quantum computing is a rapidly-emerging technology that utilizes the laws of quantum mechanics and introduces some novel quantum algorithms to solve mathematical problems that are too large or complex for classical computers \cite{AndrewSteane1998}. And quantum computers rely on qubits to run and solve these multidimensional quantum algorithms. Specifically, quantum computers take advantage of the unique behavior of quantum physics, such as superposition, entanglement, and quantum interference, and apply it to quantum computing.

%\subsubsection{Working Principle of Quantum Computers}
%Instead of bits, quantum computers use qubits. Rather than just being on or off, qubits can also be in what’s called ‘superposition’ – where they’re both on and off at the same time, or somewhere on a spectrum between the two.

%If you ask a normal computer to figure its way out of a maze, it will try every single branch in turn, ruling them all out individually until it finds the right one. A quantum computer can go down every path of the maze at once. It can hold uncertainty in its head.

%It’s a bit like keeping a finger in the pages of a choose your own adventure book. If your character dies, you can immediately choose a different path, instead of having to return to the start of the book.

%The other thing that qubits can do is called entanglement. Normally, if you flip two coins, the result of one coin toss has no bearing on the result of the other one. They’re independent. In entanglement, two particles are linked together, even if they’re physically separate. If one comes up heads, the other one will also be heads.

%It sounds like magic, and physicists still don’t fully understand how or why it works. But in the realm of quantum computing, it means that you can move information around, even if it contains uncertainty. You can take that spinning coin and use it to perform complex calculations. And if you can string together multiple qubits, you can tackle problems that would take our best computers millions of years to solve.

\subsection{Post-Quantum Cryptography} 

Quantum computers pose a serious security threat to many commonly used symmetric encryption algorithms and key negotiation schemes due to their ability to rapidly increase computational power. This increase in power heightens the risk of breaking current network security encryption schemes, leading governments and companies to transition to encryption methods in the post-quantum era. After years of research, post-quantum cryptography, which includes lattice-based, multivariate, hash-based, and code-based techniques, has emerged as one of the most developed and reliable classical computing-based approaches. 

\begin{itemize}
    \item Lattice-based cryptography: Lattice-based cryptography is a form of cryptography that utilizes the mathematical concept of a lattice, a discrete structure consisting of points connected by lines in an n-dimensional space \cite{micciancio2009lattice}.
    \item Multivariate cryptography: Multivariate cryptography is a form of cryptography that employs multivariate polynomial equations to create cryptographic primitives such as digital signature schemes and public-key encryption schemes \cite{ding2009multivariate}. 
    \item Hash-based cryptography: Hash-based cryptography relies on cryptographic hash functions to generate cryptographic primitives such as digital signature schemes and key derivation functions \cite{mozaffari2015reliable}. 
    \item Code-based cryptography: Error-correcting codes, utilized in code-based cryptography, generate cryptographic keys to secure communication \cite{overbeck2009code}. These keys have various applications, including securing communication between parties in a blockchain system.
\end{itemize}

\emph{\textbf{Lessons learned:}} The impact of quantum technology is twofold. On the one hand, quantum technology is the shield for blockchain. It achieves faster computing speed so that some complex problems can be solved in a shorter time. Meanwhile, QKD-based protocols provide higher cryptographic security. Compared with traditional computers, quantum computers reduces energy consumption and impact on the environment. On the other hand, quantum technology is the spear for blockchain. Because of the increased computational power of quantum computers, quantum computer poses a serious security threat to many classical encryption algorithms. Therefore, we need to consider the spear and shield of quantum technology and design more secure quantum blockchain solutions.

%In the context of quantum blockchain in Web 3.0, current research focuses on transitioning from classical cryptographic facilities to post-quantum cryptography. For instance, if quantum computers were to achieve a breakthrough that allows them to break classical cryptographic facilities, all blockchain systems would need to be shut down and post-quantum cryptographic algorithms installed. Until all blockchain nodes have been updated with post-quantum cryptography, the blockchain would remain offline, with decentralized transfers unable to be performed. However, this approach has two problems. Firstly, the development of quantum computers does not follow a predictable trajectory, making it difficult for network operators to accurately predict when the blockchain might need to be shut down. Secondly, the blockchain system's downtime in response to the threat of quantum computers would result in a large number of transactions being unable to be processed, leading to significant economic loss. As a result, research on quantum blockchains also focuses on how to seamlessly transition from classical cryptosystems to post-quantum cryptosystems.

\section{Motivation of Quantum Blockchain in Web 3.0}
\label{Motivation}

\begin{figure*}[!t]

\centering

\includegraphics[width=14cm]{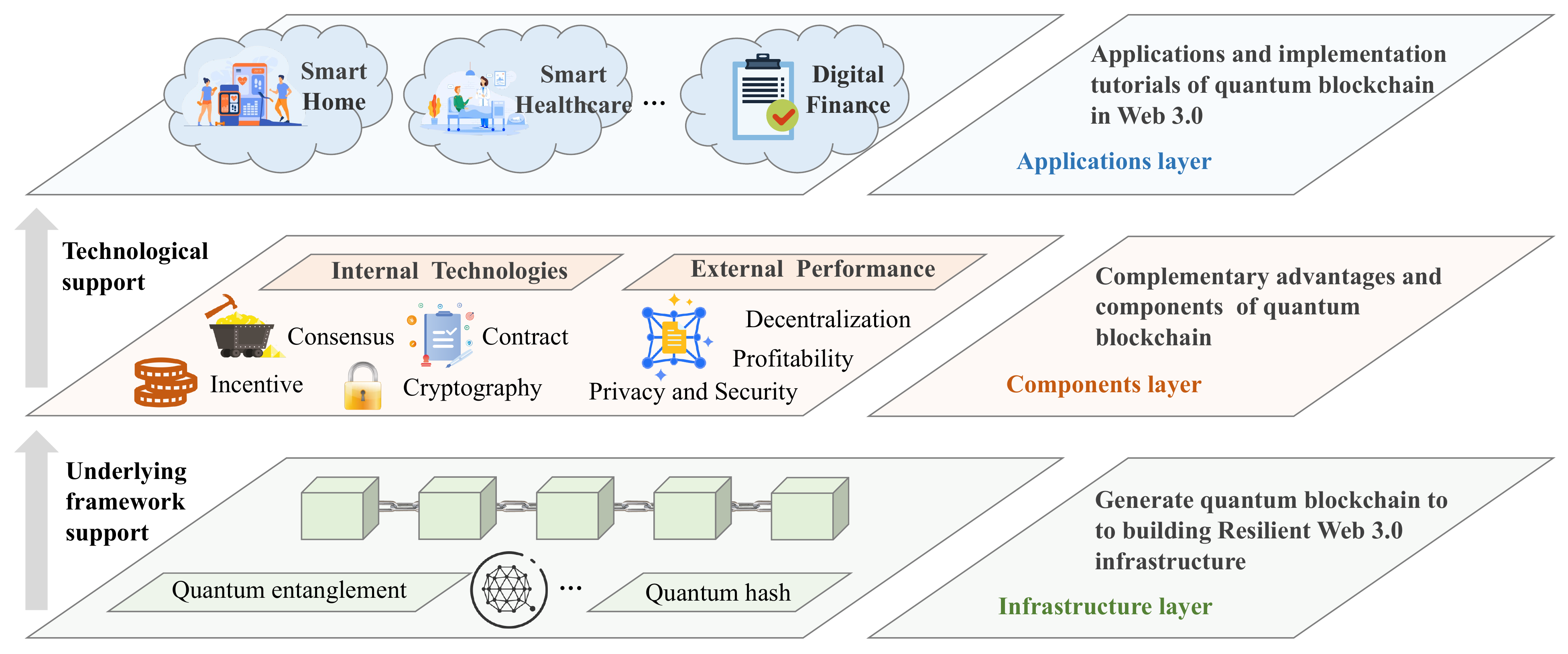}

\caption{The taxonomy graph of Quantum-Driven Blockchain in Web 3.0}

\label{fig:motivation}

\end{figure*}

The limitations of blockchain and the complementary advantages of quantum information technologies are obvious. Naturally, the emergence of quantum information technology-assisted blockchain is expected to have a significant impact on Web 3.0. In this section, we mainly discuss the motivation of quantum blockchain, as shown in Fig. \ref{fig:motivation}.
\subsection{What is the Quantum Blockchain}

Currently, there is no standard definition for quantum blockchain. Shrivas \emph{et. al} consider quantum blockchain as a distributed, decentralized and cryptographic database based on quantum information theory and computation \cite{ShrivasQB}. In \cite{Nilesh_2022}, Nilesh \emph{et. al} believes that quantum blockchains should satisfy the following properties. i) Decentralized architecture. ii) A quantum network with a distributed ledger. iii) Nodes in a quantum network need to possess quantum capabilities, e.g., quantum storage and quantum state preparation. iv) Shared quantum database. Compared with classical blockchains, quantum blockchains have the following advantages \cite{sun2018quantum}: i) More efficient. ii). More powerful. iii) More Secure. iv) Cheaper. v) Smarter. vi) Easier to regulate.

In this survey, we believe that quantum blockchain in Web 3.0 is a technology that combines the principles of quantum information technologies and blockchain, aiming to create a more secure and efficient ecosystem for Web 3.0. 

\subsection{Why Do We Need Quantum Blockchain in Web 3.0}

\subsubsection{Limitation of Blokchain} The motivation for quantum blockchains in Web 3.0 stems from the limitations of traditional blockchains, which may be vulnerable to attacks from quantum computers. Next, we describe the limitations of blockchain in two aspects.  

\textbf{Internal Technologies:} The internal compositions of blockchain face many challenges. 
\begin{itemize}
\item \emph{ Consensus protocol}, the most prominent shortcoming of consensus is the high consumption of computational resources, resulting in low scalability and long latency for Web 3.0 services. Moreover, some protocols may suffer quantum attacks. Users with quantum computers can calculate hash values, making 51\% of attacks possible.
\item  Incentive mechanism, there are certain elementary risks associated with distributed digital currencies, such as double spending, Sybil attacks, and Eclipse attacks. 
\item Smart contract, the powerful quantum computation poses serious security threats to the smart contract. Meanwhile, creating a
programming language for smart contracts in quantum logic deserves close attention and serious consideration.
\item Cryptography, with the advent of quantum computers, the main issue with classical blockchain is the possibility of cracking hash functions and classical digital signatures based on asymmetric cryptography.
\end{itemize}

 \textbf{External Performance:} The external properties face many challenges. 
 \begin{itemize}
\item \emph{ Decentralization}, the decentralized architecture is subject to security threats, such as malicious attacks, transaction tampering, etc. Although blockchain technology can provide anonymity, it may also raise privacy concerns. 
\item  \emph{Profitability}, classical cryptography in payment and voting protocols is threatened by quantum computers, raising economic issues of anonymity, verifiability, fairness, profitability, etc., in Web 3.0 scenarios.
\item \emph{Privacy and Security}, since blockchain nodes store a copy of all blocks, storage and the privacy of stored data are huge problems as the number of blocks in the blockchain continues to grow.
\end{itemize}

\subsubsection{Benefits of Quantum Information Technologies} Despite the limitations of blockchain, quantum blockchain is also motivated by the benefits of quantum information technologies. Next, we describe these benefits in two aspects.

\textbf{Internal Technologies:} Quantum information technologies bring benefits for internal compositions of blockchain. 
\begin{itemize}
\item  \emph{Consensus protocol}, the quantum consensus is more secure, efficient, and fair than traditional consensus protocols, avoiding high consumption of computational resources while providing higher throughput.
\item \emph{Incentive mechanism}, quantum blockchain provides effective mining strategies and a secure token system for incentives mechanism. 
\item  \emph{Smart contract}, on the one hand, quantum computing and quantum communication ensure the security of smart contracts. On the other hand, smart contracts based on post-quantum cryptography prevent quantum attacks.
\item  \emph{Cryptography}, some post-quantum migration methods, and anti-quantum signatures offer potential ways to achieve strong and unforgeable security under quantum attacks.
\end{itemize}

\textbf{External Performance:} Quantum information technologies bring benefits to external properties of the blockchain. 
\begin{itemize}
\item  \emph{Decentralization},  decentralized quantum blockchain tackles the issues caused by centralized management, providing a solution for achieving decentralization.
\item  \emph{Profitability}, combining classical cryptography and quantum theory, quantum cryptography has emerged to ensure the absolute security of payment protocols. Quantum technology enables anonymous, verifiable, and fair voting protocols, promoting economic prosperity and increasing profits in quantum markets.
\item \emph{Privacy and Security}, some  quantum/post-quantum encryption and verification algorithms improve the communication costs, robustness, and privacy of distributed storage.
\end{itemize}

\emph{\textbf{Lessons learned:}} Due to the complementary advantages brought by quantum information technologies, it is obvious that the converged quantum blockchain technology shows a path for implementing distributed Web 3.0 infrastructures. In the following sections, we present deployment tutorials of quantum blockchain in Web 3.0 from two aspects, including internal technologies optimization and external performance improvement.

% \subsection{Requirement of Quantum Blockchain}

\section{Internal Technologies Optimization in Quantum Blockchain }

\label{BC}

%\subsection{Non-Fungible Token}

\subsection{Quantum Consensus Protocol}

In the survey, we discuss the classification of quantum consensus protocol for Web 3.0 in detail to better understand quantum consensus. In this survey, the quantum consensus protocol refers to the improved consensus protocol in blockchain with the support of quantum technology.

\subsubsection{Quantum-Based Classification }

The quantum consensus protocols can be divided into the following two categories based on the quantum mechanical features.

\textbf{Measurement-based Consensus:} With reference to the traditional consensus mechanism of blockchain and some properties of quantum mechanics (i.e., randomness and irreversibility), a novel consensus mechanism of quantum blockchain is proposed \cite{WEN2022107693}. This consensus integrates the zero-knowledge proof and quantum measurement. Here, we present an example
to demonstrate the steps of the Quantum zero-knowledge proof protocol. David and Ben have a highly confidential secret number $A$. Without loss of generality, assume David is the side who is trying to prove that he has $A$ to Ben. The steps of the protocol are as follows.
\begin{itemize}

\item \emph{1) Distribution of shared key:} The quantum key distribution (QKD) protocol, such as BB84 protocol, is used to generate a shared key for David and Ben.

\item \emph{2) Quantum state preparation:} David prepares several pairs of EPR entanglement photons and sends another photon from each pair to Ben.
\item \emph{3) Security check:} The measurement results between two parties check the security of the quantum channel.
\item \emph{Measurement and result encoding:} The two parties perform the measurement and encode their results.
\item \emph{4) Transmission of proof information:}  David encrypts the classical bit string from encoding while sending the encryption result to Ben.
\item \emph{5) Verification of proof information:} 
By comparing the classical bit strings generated by two parties, the proof information would be verified.
\item \emph{6) Reciprocal verification:} By exchanging the roles of David and Ben, the protocol enables reciprocal verification.
\end{itemize}

Compared with the traditional consensus, the measurement-based consensus is more secure and avoids high consumption of computing resources while providing higher throughput.

\textbf{QKD-based Consensus:} Recently, QKD was proven to be unconditionally secure and has been used in quantum blockchain schemes. 
In the quantum blockchain, each pair of nodes is connected by a classical channel and a quantum channel, further forming a QKD network. Then, each pair of nodes uses the QKD network to establish a private key sequence for secure communication.

As the quantum public keys distributed by a node to other are the same, they cannot deny their actions. However, Kiktenko \emph{et. al} \cite{Kiktenko_2018} leverage QKD to distribute keys for each pair of nodes for verification. In this way, the two-way communication requires $O(n^{2})$ implementations to distribution $O(n^{2})$ keys. In particular, these works cannot provide non-repudiation because each pair of nodes shares the same key.

Based on the unconditional security provided by QKD. A quantum communication protocol is designed \cite{e21090887}. By quantum-secured communication, a set of semi-honest participants distribute the sequences of correlated numbers. These participants then exchange some information to reach a consensus. In the further development of this protocol, low dimensional entanglement can be considered to replace the key distribution.

\subsubsection{ Blockchain-Based Classification}

Referring to \cite{Nguyen18survey}, consensus protocols in blockchain are classified into two groups in Web 3.0. The first group is the proof-based consensus protocol that uses computational proofs to determine the winner-generating block, such as PoW. The second group is the vote-based consensus protocol. In this regard, the nodes in the blockchain have equal votes and reach a consensus, such as PBFT. According to the blockchain-based classification, we classify the quantum consensus protocol in the same way.

\textbf{Proof-based Consensus:} Traditional proof-based consensus in Web 3.0 is computationally intensive and time-consuming, making blockchain systems inefficient. 

To improve the blockchain consensus, the post-quantum threshold signature scheme is exploited \cite{YI2021100268}. This scheme is to solve quadratic equations in a finite field, which is resistant to attacks by quantum computers. Based on this novel method, more than one and a half nodes in the blockchain network sign the new block using the post-quantum threshold signature. As a result, it has a higher level of security. Furthermore, much of the computational complexity of the consensus comes from the signature of the new block, which is much more efficient than the existing consensuses.

Compared to current threshold signature schemes (e.g., RSA-based and elliptic curve signature-based algorithms), post-quantum threshold signatures are more promising and advantageous for future blockchain and other applications.

\textbf{Vote-based Consensus:} Due to the node-independent computing power, the quantum capabilities of all parties are unbalanced, causing proof-based consensus to be inapplicable. Therefore, the vote-based consensus is more suitable for quantum blockchain in Web 3.0.

\begin{figure*}[!t]

\centering

\includegraphics[width=10cm]{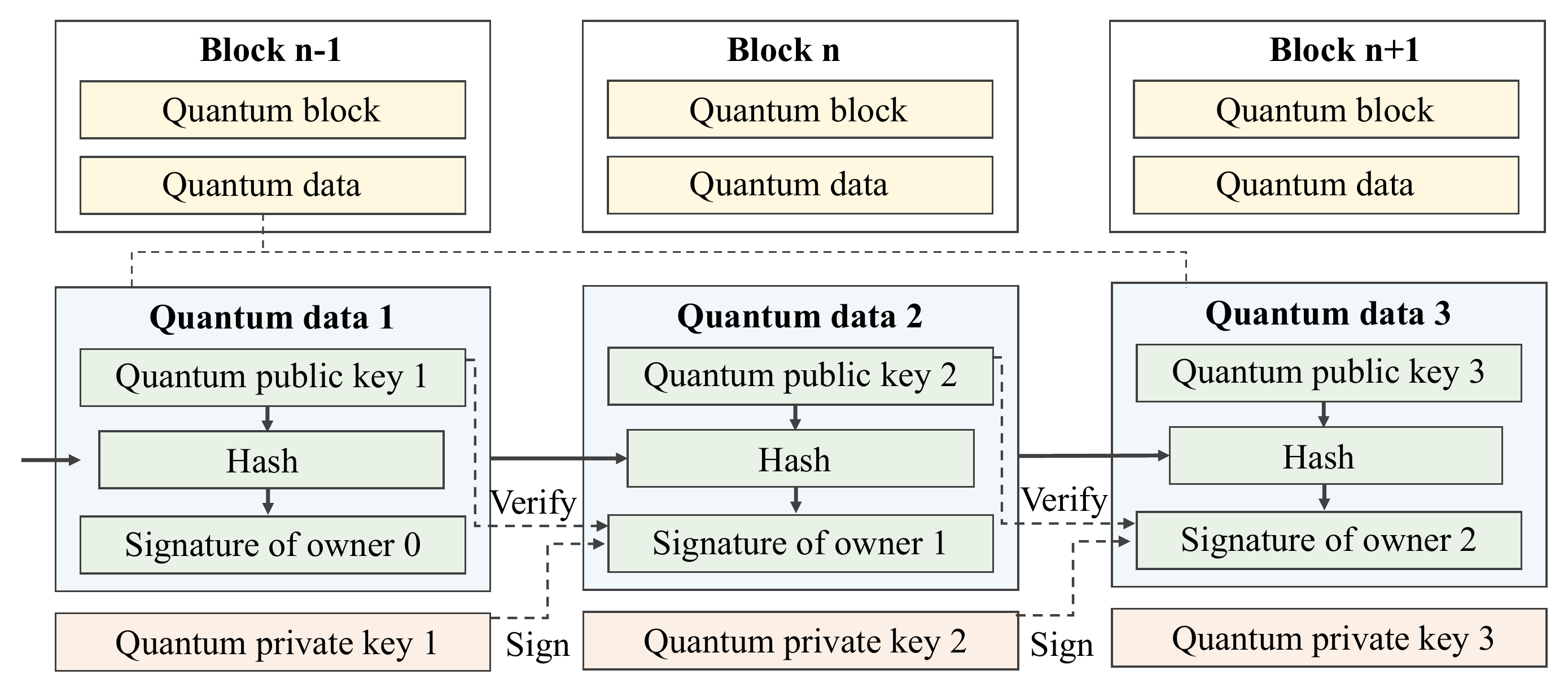}

\caption{The structure of quantum blockchain}

\label{fig:quantum_blockchain}

\end{figure*}

To ensure the fairness of vote-based consensus, a quantum delegated proof of stake (QDPoS) based on quantum voting is constructed \cite{QDPoS222Wu}, allowing for fast decentralization even if the quantum computer emerges in the future. Like DPoS, QDPoS elects a certain number of representative nodes to generate new blocks by quantum voting. The quantum digital signature is introduced to ensure the quantum blockchain's efficiency and security. Based on QDPoS and quantum digital signature, a quantum blockchain scheme is designed. Its structure is shown in Fig. \ref{fig:quantum_blockchain}. Notably, quantum blockchain uses quantum states to construct quantum blocks connecting by leveraging the entanglement property between quantum states. 

As an extensively-applied consensus, PBFT is derived from the Byzantine Generals' problem. To achieve Byzantine agreement between multiple parties, a quantum byzantine agreement without entanglement is proposed \cite{Sun_2020Multi}. The unconditional security of quantum key distribution contributes to the distribution of the relevant digital sequences, ensuring the security of the Byzantine agreement protocol. To improve the efficiency of consensus,
a quantum honest-success Byzantine agreement protocol introduces quantum protection, and quantum certificate into the syntax of transactions \cite{Quantum18Sun}. Similar to the work in \cite{Sun_2020Multi}, there is no use of multi-particle entanglement in these quantum blockchains, making them easy to implement with existing technology.

%\subsubsection{ Implementation Tutorial for Quantum Consensus Protocol}

\subsection{Incentive Mechanism}

Through incentive mechanisms are well-studied in classical blockchain~\cite{han2022can}, the problem of how to design effective incentive mechanisms for participants in the quantum blockchain is still in its fancy in Web 3.0. To maintain the decentralization of classical blockchain, incentives, e.g., static mining rewards and transaction gases, are distributed for motivating miners to participate during the consensus process. 

\subsubsection{Quantum-based Mining Strategies} In addition to the PoW and PoS in classical blockchain in Web 3.0, the miners in a quantum blockchain leverage quantum resources, e.g., QKD, to record new transactions and blocks. Then, minors are rewarded with tokens in the quantum blockchain. For example, Azhar \textit{et al.}~\cite{azhar2019blockchain} use quantum cryptography protocols such as QKD to guarantee that communications in blockchain between participating nodes can be secure from third-party intrusion. Through the use of the six-stage QKD protocol~\cite{coles2016numerical}, the transparency and immunity of blockchain-based cryptocurrency systems can be ensured. Furthermore, by observing the key rate generation on the NumericalQKD simulation platform, the authors experimentally verify the feasibility of the protocol. This study provides a direction for future quantum blockchain research. The above scenario for the key generation protocol can be implemented when any blockchain system is developed.

Based on the need to set up additional infrastructures such as quantum servers, the authors in~\cite{bennet2019energy} propose the concept of proof-of-entanglement (PoE), which allows mining nodes of a quantum blockchain to reach consensus based on quantum resources. Specifically, an energy-efficient interactive protocol is proposed that allows quantum blockchain miners to execute formulas based only on their optically encoded quantum information without having to trust the network infrastructure.

\subsubsection{Secure Token Systems} Similar to the idea of reaching a consensus based on entanglement resources, Sun \textit{et al.} in~\cite{sun2018quantum} propose the incentive mechanism, namely Qulogicoin, for the quantum-enhanced logic-based blockchain. In this mechanism, the Byzantine agreement protocol is used with honest success to maintain the quantum blockchain under various fraudulent attackers, such as double-donating adversaries, manipulating adversaries, and bribed distributors. Through implementation and testing based on current technologies, the authors demonstrate that the proposed incentive mechanism is efficient and powerful. In addition to the fungible tokens, non-fungible tokens (NFTs) for quantum blockchain integrates the best features of blockchain technology to provide a unique and authentic token, each with unique properties and non-substitutable resources. Unfortunately, in terms of energy consumption for mining and lack of security, the current classical NFTs are costly. Therefore, instead of making quantum states representing NFTs physically available to their owners, Pandey \textit{et al.} in~\cite{Pandey22NFT} propose a new protocol for preparing quantum irreplaceable tokens. The protocol is ultimately more efficient than Proof of Work (PoW) and incorporates Proof of Stake (PoS). The proposed scheme is simulated and analyzed against various quantum computing attacks, including intercept and retransmit, entangle and measure, man-in-the-middle, and classical threats. It demonstrates its ability to withstand these attacks. The authors expect the proposed protocol to replace classical NFTs as quantum hardware evolves. These quantum NFTs can be used to store patient monitoring and medical data cheaply and securely in the quantum blockchain. Kaushik \textit{el al.} in~\cite{kaushik2022demystifying} propose an incentive mechanism for storing health data in the quantum blockchain that can ensure veracity, integrity, and availability.

% Blockchain based Secure Crypto-currency system with quantum key distribution protocol~\cite{azhar2019blockchain}

\subsection{Smart Contract}

Existing smart contracts in Web 3.0 against quantum attacks strive to provide solutions for efficiency and scalability. On the one hand, quantum computing and quantum communication can provide demonstrable security in the development of smart contracts. On the other hand, smart contracts based on post-quantum cryptography are partially secure against quantum attacks. These two solutions can provide efficient and scalable solutions for signing, executing, and verifying smart contracts.

\subsubsection{Quantum Cryptography-based Smart Contracts} The authors in~\cite{cai2019blockchain} propose a smart contract-based architecture to defend against quantum attacks in quantum networks, based on quantum blind signatures. The theoretical analysis of the authors shows that the quantum signature scheme proposed in this paper, which is based on the properties of quantum entanglement, can be used for both single and multiple signatures, and can improve the security of blockchain smart contracts against quantum attacks. First, the authors carefully analyze the message processing and transmission of the quantum blind signature in this framework. Then, they describe the lifecycle and signature rules of quantum-blind signatures for smart contracts. In addition, the authors explain the algorithm design and operation, analyze the algorithm's security performance, and propose a quantum blind signature protocol for a lightweight signature of smart contracts. Finally, based on the original single signature algorithm, a more complex quantum blind signature algorithm for multiple signatures is proposed, and its security performance is analyzed. Furthermore, Cai \textit{et al.} in 
\cite{cai2021quantum} propose a framework for multiparty transactions based on the quantum blind multi-signature method in a decentralized manner. In this framework, the authors develop a quantum blind multi-signature algorithm that includes initialization, signing, verification, and implementation to provide computationally efficient and scalable solutions for the quantum blockchain.

\subsubsection{Post-quantum Smart Contracts} For a blockchain relying on smart contracts, the usefulness of smart contracts against quantum attacks can also be strengthened by post-quantum cryptography. For example, Karbasi \textit{et al.} in~\cite{karbasi2020post} have proposed an end-to-end encryption scheme to protect against man-in-the-middle and eavesdropping attacks. 
In \cite{sun2021logic}, to use a logic-based smart contract programming language in a recently proposed quantum-secure blockchain framework, the authors investigate the use of logic and logic programming in smart contract design. The authors propose a logic-based smart contract programming language called Logicontract (LC), where the logic used in LC is extended with modern declarative logic programming methods. The introduction of post-quantum signatures overcomes the specific limitations of LC, where unconditionally secure signatures, despite their strength, have limited protection for users of the same node. The next step is to investigate its use for secure multiparty computing. Another goal is to extend our logic language to include the original quantum language, thus creating a programming language for smart contracts in quantum logic. As an important application in cloud computing, Li and others in \cite{li2023post} are exploring the use of post-quantum blockchain for provable data ownership. To avoid the single point of failure and partial trust caused by a centralized external auditor. Grid-based, privacy-preserving, verifiable data ownership is enabled based on smart contracts. The analysis demonstrates the correctness, soundness, and performance of the proposed system through a series of interactive games.

\subsection{Post-Quantum Migration for Blockchain}

To effectively avoid the risk associated with pre-quantum blockchain, some post-quantum migration methods have been proposed by many researchers in recent years. In this subsection, we explore the implementation of post-quantum migration in the following mechanisms. We list details of the existing research works in Table \ref{tab:migration}. 
\begin{table*}		
	\renewcommand{\arraystretch}{2}	
	\centering		
	\caption{Details of the existing post-quantum migration methods for blockchchain in Web 3.0.}		\label{tab:migration}
	\fontsize{7.5}{7.5}\selectfont	
	\begin{tabular}{m{1.8cm}<{\centering}|m{1cm}<{\centering}m{2.5cm}<{\centering}m{2cm}<{\centering}m{6cm}}		
		\hline	
            \hline
		\textbf{Types} &\textbf{Ref. }  & \textbf{Required Technology} & \textbf{Benefits} &\makecell[c]{\textbf{Contribution}}\cr 		
		\hline		
		\multirow{6}{*}{ \textbf{	\rotatebox{90}{\tabincell{c}{Public key post  \\ quantum cryptosystems }}}} &
		\cite{alagic2019status,alagic2020status} ~&Post-quantum cryptography& Security & Adjust the required security parameters of the cryptosystem
		\\
	\cline{2-5}		&\cite{ding2004new,ding2005cryptanalysis} ~& Square matrices with random quadratic polynomials& Higher decryption speed & Propose a new variant of the cryptosystem which is more suitable for Web 3.0 scenario
		\\
		\cline{2-5}			&\cite{hoffstein2006ntru} ~&Polynomial algebra &  Security, efficiency & Propose a new public key cryptosystem, allowing for accelerated blockchain user transactions 
            \\
          \cline{2-5}	
	&\cite{alkim2016post,GoogleHybrid} ~&Key-exchange protocol& Post-quantum security & Propose new parameters and a better suited error distribution to protect the exchanged data from quantum attacks
  		\\
		\hline
  		\multirow{8}{*}{ \textbf{	\rotatebox{90}{\tabincell{c}{Post-Quantum \\Signature Algorithms  }}}} &
		\cite{singh2020securing,gottesman2001quantum} &Quantum digital signatures, quantum teleportation &Security, privacy &  Propose secure mechanisms for performing transactions on the blockchain network \\
	\cline{2-5}	
	&	\cite{gao2018secure} &Quantum-resisting signature &Low computational complexity, unforgeability &  Present post-quantum blockchain and a secure cryptocurrency scheme to resist quantum computing attacks \\
 	\cline{2-5}	
	&	\cite{yin2018anti,li2018new} & Anti-quantum signature &Lightweight, security &  Propose an anti-quantum transaction authentication scheme in the blockchain\\
 	\cline{2-5}	
	&	\cite{guneysu2012practical} &Post-Quantum Cryptography &Practicability &  Develop a signature scheme for practicability and use in embedded systems \\
 	\cline{2-5}	
	&	\cite{an2018qchain} &Post-quantum cryptography &Internal technologies &  Introduce a modified lattice-based signature scheme for decentralized PKI system \\
		\hline
  		\multirow{1}{*}{ \textbf{	\rotatebox{90}{\tabincell{c}{Quantum  \\Networked\\ Time Machine  }}}} &		\cite{megidish2012resource} & Quantum entanglement& Scalability & Present a new approach generating quantum entanglement between many photons by using only a single source of entangled photon pairs to solve the scalability problem \\
	\cline{2-5}	
	&	\cite{quantum1010002} &Quantum entanglement  & Security &  Overview of a conceptual design of a quantum blockchain using entanglement, which includes encoding the blockchain as a temporal Greenberg-Horn-Chalinger state of photons\\
  			\hline
  		\multirow{3}{*}{ \textbf{	\rotatebox{90}{\tabincell{c}{Quantum \\Hashing  }}}} &
		\cite{jin2009quantum} &Quantum hash &Robustness & Present a novel multimedia identification system based on quantum hashing \\
	\cline{2-5}	
	&	\cite{swaminathan2006robust} &Fourier transform features, controlled randomization &Security, robustness &  Develop a novel algorithm for generating an image hash  \\
		\hline
  \hline
	\end{tabular}	
\end{table*}

\subsubsection{Public Key Post Quantum Cryptosystems }

There exist four types of post-quantum cryptosystems and a fifth type of cryptosystem with a mixture of pre-quantum and post-quantum.

\textbf{Code-based cryptosystem:} It is based on the theory supporting error-correcting codes. For example, McEliece's cryptosystem can perform fast encryption and decryption in blockchain \cite{mceliece1978public}. Nonetheless, this cryptosystem requires storing and executing operations of large matrices, a restriction for resource-constrained devices. To address this problem, some code-based post-quantum encryption cryptosystems investigate matrix compression techniques, as well as specific coding
techniques. Notably, the NIST call for post-quantum public-key cryptosystems is currently in its second round \cite{alagic2019status,alagic2020status}, with the first standard draft expected to be released between 2022 and 2024. These cryptosystems adjust the required security parameters while ensuring classical security between 128 and 256 bits. Compared with the current ECDSA \cite{koblitz1987elliptic} and RSA-based \cite{rivest1978method} encryption systems,  these post-quantum encryption cryptosystems offer the best trade-off between security and key size.

\textbf{Multivariate-based cryptosystem:} It depends on the complexity of solving systems of multivariate equations, which turns out to be NP-hard \cite{petzoldt2010selecting}. 

Currently, some multivariate-based schemes are based on the use of square matrices with random quadratic polynomials, cryptographic systems derived from the Matsumoto-Imai algorithm \cite{ding2004new}, and schemes depending on hidden field equations \cite{ding2005cryptanalysis}. These schemes are more suitable for Web 3.0 scenarios by reducing the key size and ciphertext overhead while increasing the decryption speed.

\textbf{Lattice-based cryptosystem:} It depends on the presumed hardness of lattice problems \cite{blomer2009sampling} that cannot currently be solved by quantum computers.

Lattice-based schemes allow for accelerated blockchain user transactions and can be executed efficiently as they are generally simple to calculate. The implementations of these lattice-based schemes require the storage and utilization of large keys \cite{hoffstein2006ntru}. The general schemes guarantee classical security between 128 and 368 bits and quantum security between 84 and 300 bits. Based on the security level they provide, the complexity of these schemes varies significantly.

\textbf{Hybrid cryptosystem:} It merges pre-quantum and post-quantum cryptosystems with the aim of protecting the exchanged data from quantum attacks. For instance, X25519 merging New Hope \cite{alkim2016post} with an ECC-based Diffie-Hellman key agreement scheme has completed the testing by Google \cite{GoogleHybrid}.

Different from the existing public key infrastructure, the quantum state information cannot be copied. Thus, only those who know the exact quantum state can create and distribute public keys. Additionally, all the above cryptosystem implementations require significant computational resources and more energy consumption. Therefore, the post-quantum cryptosystems for blockchain should aim to find a balance between security, computational complexity, and resource consumption.

\subsubsection{Post-Quantum Signature Algorithms}
The rapid development of quantum computers poses significant security risks to the cryptographic algorithms underlying blockchain networks. To overcome the vulnerabilities of the blockchain networks to quantum adversaries, some potential anti-quantum signature
algorithms have been proposed. 

Using quantum digital signatures and quantum teleportation phenomenon, a quantum secure theme is presented to implement the blockchain \cite{singh2020securing}. In particular, messages are signed by quantum digital signatures \cite{gottesman2001quantum} and then they are distributed across the blockchain network by the teleportation phenomenon, creating the opportunity to build a more secure and faster blockchain for Web 3.0.

Based on lattice cryptography, novel quantum-resisting signature schemes are used for transaction authentication in blockchain-enabled systems. In particular, the lattice basis delegation algorithm provides a way to generate secret keys \cite{gao2018secure}. Then, the double signature, defined as the first signature and last signature, is developed to reduce the correlation between the message and the signature. The secure cryptocurrency scheme combines the proposed signature scheme with the blockchain, satisfying correctness and having the advantage of resisting quantum computing attacks.

Different from the previous elliptic curve signatures, a new signature authentication
scheme leverage Bonsai Trees technology to generate the extended lattice accompanied by the corresponding key and then generate the signature \cite{yin2018anti}. Using this signature, the anti-quantum transaction authentication approach achieves strong and unforgeable security under selected message attacks. In addition, by combining Bonsai Trees technology and the RandBasis algorithm, public and private keys can be generated for verifying the transaction message \cite{li2018new}. Meanwhile, the size of the public key, private key, and signature are smaller than the similar methods, further decreasing the computational complexity and increasing the implementation efficiency. Regarding the work in \cite{guneysu2012practical}, a lattice-based signature scheme is developed to optimize embedded systems. For a 100-bit security level, 9,000 bits signatures need to be generated by using a 12,000-bit public key and a 2,000-bit private key. Due to its simplicity and efficiency, this scheme is considered a new signature algorithm for managing public key encryption for a post-quantum decentralized system, such as QChain \cite{an2018qchain}.

%However, some researchers consider XMSS and SPHINCS
%to be impractical for blockchain applications due to their
%performance [202], so alternatives have been suggested. For
%example, XMSS has been adapted to blockchain by mak-
%ing use of a single authentication path instead of a tree,
%while using one-time and limited keys in order to pre-
%serve anonymity and minimize user tracking [203].
\subsubsection{Quantum Networked Time Machine} It is a conceptual design
for quantum blockchain. This means if a quantum blockchain are to be constructed, it could be regarded as a quantum networked time machine.

Entanglement in time, as opposed to entanglement in space, plays a key role in the quantum benefits of classical blockchains. In \cite{megidish2012resource}, the authors use the temporal GHZ state of the photon as a blockchain, providing a key quantum advantage over spatial entanglement. In this conceptual system, a temporal Bell state as a block contains two classical records, and a growing temporal GHZ state has been taken as a chain.

In this way, the functionality of timestamped blocks and hash functions are replaced with temporal entanglement \cite{quantum1010002}. The quantum advantage is a significant amplification of the susceptibility to tampering, meaning that if one user tampers with a single block, the complete local copy of the blockchain might be destroyed. While for a classical blockchain, only the block after the tampered block is destroyed, making it vulnerable to vulnerabilities. This work offers the realistic possibility of deploying pure quantum blockchains.

\subsubsection{Quantum Hashing} 

It aims to improve the robustness of the binary hashing systems using the same intermediate hash values against various distortions.

Based on the severity of the distortion, the Hamming weight of the D-bit binary hash difference between a query and its true-underlying content can vary from 0 to D. The probability that each bit of the D-bit binary hash difference evaluated by the quantum hashing system is 1. Instead of using a definite hash value in a binary hash, quantum hashing incorporates uncertainty in the hash value \cite{jin2009quantum}. More importantly, the main difference between a quantum hashing system and a binary hashing system is the way the dissimilarity between the query and the candidate data in the binary hash database is calculated.

In the binary hashing system, the system extracts the intermediate hash from the query. It encodes it as a binary hash to find the matching pair in the binary hash database \cite{swaminathan2006robust}. Then, the dissimilarity between the query and the candidate data is obtained. Compared to this binary hashing system, the quantum hashing system does not require additional storage as the quantum hashing algorithm uses the same binary hash database as the binary hashing algorithm. Moreover, acceptance and rejection decisions can be made with a slight increase in computational cost.
It provides a novel post-quantum migration algorithm for blockchain against quantum attacks.

\emph{\textbf{Lessons learned:}} In the survey, we explore the internal technologies optimization in the quantum blockchain. It aims to improve the internal components of the blockchain, including designing unconditionally secure quantum consensus, effective incentive mechanisms, and scalable smart contracts. effective incentive mechanisms, and scalable smart contracts. These show that quantum information technologies offer promising solutions for blockchain optimization that can further facilitate the construction of Web 3.0 infrastructure.

%\subsubsection{Implementation Tutorial for Post Quantum Cryptography}

\section{External Performance Improvement of Quantum-Driven Web 3.0}

\label{Edge}

\subsection{Quantum Blockchain Economy}

The quantum blockchain economy refers to establishing a secure and decentralized Web 3.0 economic model with the support of quantum and blockchain technology. In this section, we mainly explore the quantum blockchain economy and list details of the existing research works in Table \ref{tab:external}. 

\subsubsection{Payment Protocol} 
E-commerce has become a major force in the global economy because of the convenience and accessibility it provides. The growth of e-commerce has also led to the development of new payment models, such as electronic cash (E-cash) payment protocols. Compared with other protocols, E-cash is becoming a more desirable payment protocol due to its anonymity and offline transferability.
Currently, the security and cryptography of electronic payment have been the focus of research. Particularly, classical cryptography in electronic payment might be threatened by quantum computers with the development of quantum computing \cite{grover1996fast}.

To address the above challenges, quantum cryptography, combining classical cryptography and quantum theory, appears to ensure the unconditional security of payment protocol. Based on the quantum blind group signature with two trusted third parties \cite{xiaojun2010payment}, an E-payment system is established, and then a cross-bank electronic payment protocol based on the quantum proxy blind signature emerges \cite{wen2013inter}. Nevertheless, the presence of a dishonest customer in a payment protocol leads to a risk that some purchase information may be leaked. Different from the proxy blind signature, quantum multi-proxy
blind signature and group blind signature can be used in the e-payment protocol with a third-party platform to further guarantee the customer’s anonymity \cite{shao2017payment,zhang2017third}. 

Blockchain acts as a trusted third party, integrated with quantum technology, to achieve reliable trust between both parties of the transaction \cite{gou2021novel,zhang2019novel}. These studies propose an e-payment protocol implemented by blockchain and quantum signature. Specifically, these works describe in detail the main characters and implementation process of the quantum e-payment system as follows.

\textbf{Main characters:} i) Customer. (ii) Two customer agent banks, denoted as bank $A$ and bank $B$, respectively. (iii)Representative of the bank, denoted as $R$. (iv) Merchant. (v) Blockchain.

\textbf{Implementation process:} 
When a customer pays for online purchases, he finds that bank A has a low balance. In this case, he runs out of bank A's money, and then he pays bank B the rest of the price. In this process of transaction, $R$ receives the purchase information of the customer by QKD protocols. Meanwhile, $R$ plays as the proxy signer to sign the payment slip when the settlement is completed. After, the customer sends a signed blind copy to the merchant by using the Bell-state measurement performed by agent banks $A$ and $B$. 
%The brief steps of this scheme are described in Fig. 2.

By leveraging the property of quantum proxy blind signature, the anonymity of e-payment is guaranteed. Furthermore, blockchain technology stores all the information of customers and helps the merchant verify the legitimacy of the signature. This scheme has good application prospects and wide application value.

\subsubsection{Market Mechanism}
A number of voting protocols based on quantum blockchain have been developed to address the risks posed by the upcoming quantum computers, providing anonymous, verifiable, fair, and self-tallying voting while simplifying the task of electronic voting. Similar to the voting protocol on the Bitcoin blockchain, the voting protocol supported by the quantum blockchain consists of two phases: a ballot commitment phase and a ballot tallying phase \cite{sun2019simple}. In this voting protocol, quantum secure communication (QSC) is leveraged to distribute the matrix by voters, then voters give the masked ballots to miners. Afterward, quantum byzantine agreement (QBA) is used by miners to achieve consensus about voters’ masked ballot. 
%The visualization of this protocol is shown in Fig. \ref{}.
\begin{figure}[!t]

\centering

\includegraphics[width=7cm]{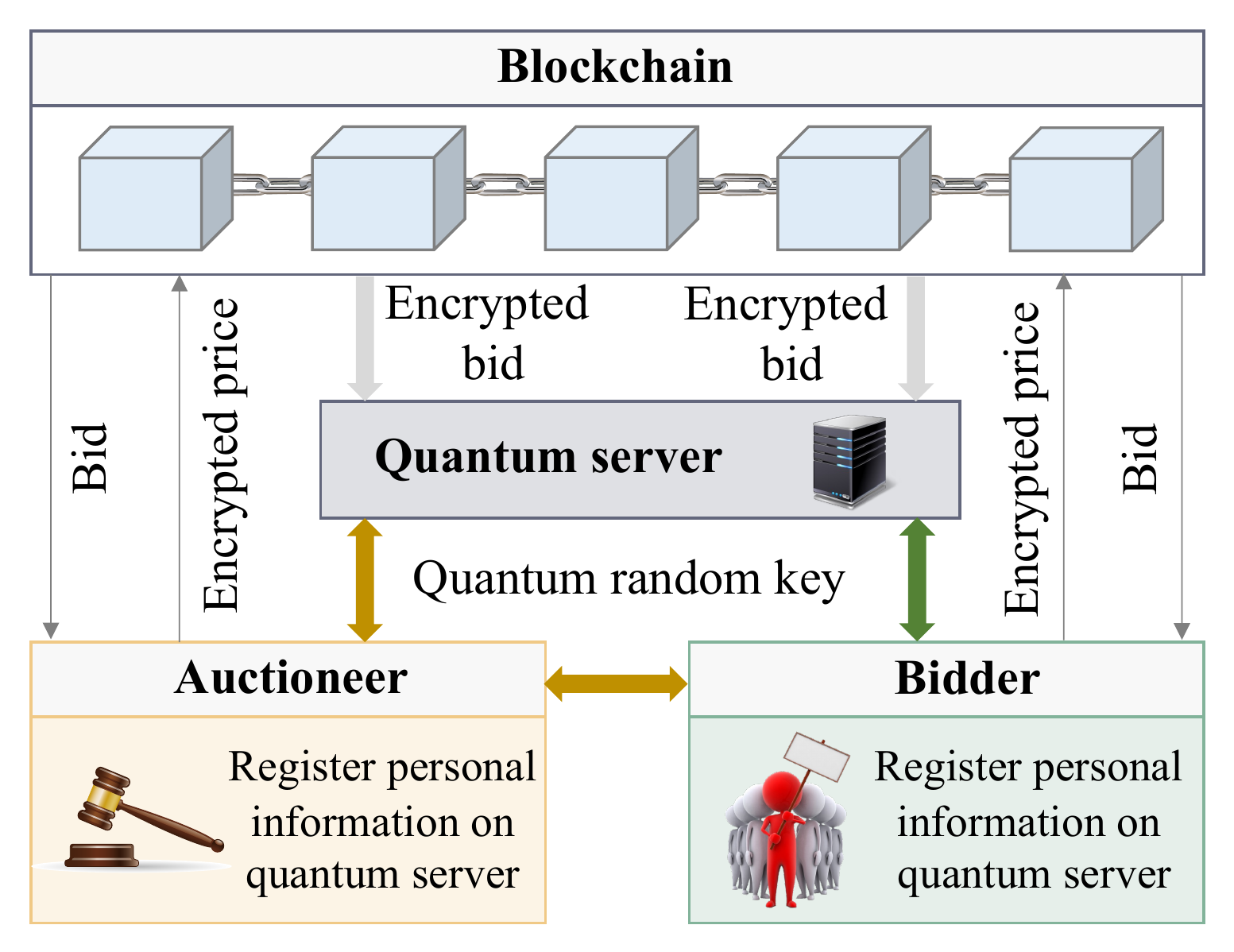}

\caption{An auction on the quantum blockchain}

\label{fig:auction}

\end{figure}

Moreover, auctions are one of the more important market mechanisms in the quantum blockchain economy. It allocates resources according to a set of market rules and prices determined by buyers' bids. Blockchain-based auctions \cite{blass2018strain} satisfy decentralization in auctions, and quantum auctions \cite{zhang2018economic,liu2016multiparty} satisfy unconditional security, but existing auction protocols cannot guarantee both properties. In \cite{sun2020lottery}, an auction on the quantum blockchain is proposed to solve this problem. Fig. \ref{fig:auction} shows a simple flow of the auction process. There are three types of participants in this auction protocol, sellers, buyers, and miners. This auction process is divided into five stages,
\begin{itemize}
\item \emph{1) The bidding phase:} Each buyer announces his bid to the sellers and to all miners by quantum bit Commitment.
\item \emph{2) The opening phase:} Each buyer opens his bid to the sellers.
\item \emph{3) Decision phase:} The winning bid and the winning buyer are determined.
\item \emph{4) Verification phase:} This stage convinces miners that the buyer has picked an effective winner.
\item \emph{5) Publication phase:} Achieving consensus and adding it to the blockchain.
\end{itemize}

In general, the transactions of the auction are stored in the blockchain and supported by quantum computation and communication to enhance security and protect privacy. As one of the major auction protocols, the sealed-bid auction protocol using a quantum-based blockchain is explored \cite{abulkasim2021quantum}. This protocol uses quantum computing and blockchain technology to guarantee essential features and requirements. 
%The visualization of this protocol is shown in Fig. \ref{}.
\begin{itemize}
\item \emph{1) Registration:} Bidders interested in the auction register their personal information on the quantum server.
\item \emph{2) Generating quantum keys:} Quantum server shares random QKD-based keys with the auctioneer and bidder to encrypt data and authenticate their identities. 
\item \emph{3) Data encryption and submission:} Each bidder and the auctioneer encrypt the private bid and the lowest acceptable price using the encryption keys, and submit the encrypted data to the blockchain, respectively. 
\item \emph{4) Data evaluation and announcing the winning bid:} The quantum server downloads all submitted bids and the auctioneer's encrypted data to calculate the winning bids and submit them to the blockchain.
\item \emph{5) Verification:} Each bidder verifies the winning bid by comparing the submitted bid with the winning bid.
\end{itemize}

The quantum blockchain-based lottery and auction protocols satisfy the significant features of distributed lotteries and auctions and can be implemented by existing technologies. It is believed that quantum blockchains can provide new insights into its market economic mechanisms.

\begin{table*}		
	\renewcommand{\arraystretch}{2}	
	\centering		
	\caption{Details of the existing works in external performance improvement.}		\label{tab:external}
	\fontsize{7.5}{7.5}\selectfont	
	\begin{tabular}{m{1.2cm}<{\centering}|m{0.8cm}<{\centering}m{2cm}<{\centering}m{2cm}<{\centering}m{4cm}m{1.1cm}<{\centering}m{1.1cm}<{\centering}}		
		\hline	
            \hline
		\textbf{Type} &\textbf{Ref. }  & \textbf{Quantum Technology} & \textbf{Benefits from quantum} &\makecell[c]{\textbf{Contribution}}&\textbf{Security Analysis}&\textbf{Storage Analysis}\cr 		
		\hline		
		\multirow{10}{*}{ \textbf{	\rotatebox{90}{\tabincell{c}{Quantum Blockchain  \\Economy }}}} &
		\cite{xiaojun2010payment} ~&quantum key
distribution& Unconditional security & Propose a new E-payment system based on quantum blind and group signature & \checkmark&  $\times$
		\\
	\cline{2-7}		&\cite{wen2013inter,
 shao2017payment,zhang2017third} ~&Quantum proxy blind signature& Security, anonymity&Develop e-payment systems to support inter-bank transactions & \checkmark&  $\times$
		\\
		\cline{2-7}			&\cite{gou2021novel,zhang2019novel} ~&Quantum proxy blind signature, quantum key distribution&Security, anonymity& Describe in detail the implementation process of the quantum
e-payment system  & \checkmark&  $\times$
            \\
          \cline{2-7}		
	&\cite{sun2019simple} ~&Quantum computation& Anonymity, eligibility & Develop a simple voting protocol based on quantum blockchain & \checkmark&  $\times$
  		\\
		\cline{2-7}			
&\cite{blass2018strain,zhang2018economic,liu2016multiparty}  &Quantum communication & Security, fairness &  Present the economic and feasible quantum auction protocols& \checkmark&  $\times$ \\ 
		\cline{2-7}			
&\cite{sun2020lottery,abulkasim2021quantum}  & Quantum bit commitment & Unpredictability, unforgeability, verifiability, decentralization  &  Present a protocol for auction on quantum blockchain& \checkmark&  $\times$ \\ 
		\hline
  		\multirow{-1.2}{*}{ \textbf{	\rotatebox{90}{\tabincell{c}{Decentralization  }}}} &
		\cite{yang2022decentralization} &Quantum computing, quantum networks & Decentralization, security &  Provide a theoretical analysis of the quantum blockchain for decentralized identity authentication& \checkmark&  $\times$\\
		\cline{2-7}		
	&	\cite{wang2009theoretical} &Quantum identity authentication, zero-knowledge proof & Unconditional security & Propose a theoretical scheme for zero-knowledge proof quantum identity authentication & \checkmark&  $\times$\\
 		\hline
  		\multirow{2}{*}{ \textbf{	\rotatebox{90}{\tabincell{c}{Scalability  }}}} &	\cite{coladangelo2020quantum} &Quantum networks & Decentralization, scalability &  Integrate smart contracts and quantum lightning to improve the speed of payment transactions & \checkmark&  $\times$\\
		\cline{2-7}		
	&	\cite{bhavin2021blockchain} &Quantum teleportation &  Low throughput and latency & Propose a consensus mechanism to reduce resource consumption &  $\times$& \checkmark\\
		\hline
  		\multirow{4.5}{*}{ \textbf{	\rotatebox{90}{\tabincell{c}{Privacy of \\ storage  }}}} &
		\cite{mesnager2020threshold} &Post-quantum encryption &Low storage cost, robustness &  Propose a secure threshold verifiable multi-secret sharing scheme for transaction verification and private communication & \checkmark&  \checkmark\\
		\cline{2-7}		
	&	\cite{raman2018distributed,kim2018efficient} &Quantum teleportation &  Low throughput and latency & Present a secure threshold-based verifiable multi-secret while improving the
recovery communication cost and robustness  &  $\times$& \checkmark\\  	
		\cline{2-7}		
	&	\cite{chen2022aq} &Quantum signature &  Anonymity, unforgeability, shareability & Develop an anti-quantum signature for secure data sharing with blockchain  & \checkmark& $\times$ \\  
		\hline
  \hline
	\end{tabular}	
\end{table*}

\subsection{Decentralization}

In the quantum blockchain, classical data chains are redesigned in a quantum way by associating quantum states with entanglement. In this case, quantum decentralization is claimed to be ``unconditionally secure". Moreover, quantum blockchain infrastructure, such as access control and user authentication, must be established prior to any quantum decentralization. In this survey, we focus on decentralized identity management in the quantum blockchain.

The current Internet keeps the digital identities of users in centralized storage. Centralized identity violates user privacy and may experience problems such as single points of failure or information tampering. Decentralized identity management based on quantum blockchain has been proposed to tackle the issues caused by centralized identity. The quantum blockchain identity framework (QBIF) is a popular method for implementing secure pseudonyms \cite{yang2022decentralization}. In QBIF, users control their identities and use them without revealing unnecessary information. Meanwhile, identities are secure and authentic in a decentralized quantum environment, further protecting privacy and preventing forgery \cite{al2017scpki,pinter2019towards}. Next, we mainly introduce various aspects of the QBIF architecture, including roles, quantum identity attestations (QIAs), and quantum blockchain.

\subsubsection{Roles in QBIF} 
Within this framework, QBIF has three roles, including identity owners, identity issuers, and identity verifiers. Identity issuers represent trusted parties that issue identities. In addition, they verify the validity of identity attributes and issue attestations, stored on blockchains. 
Users, i.e. owners, determine their own identity and use on-chain authentication to prove to verifiers their identity. The verifiers provide services based on the reputation of the issuer who offers the attested identity.

\subsubsection{Quantum Identity Attestations} 
Quantum identity attestations, also called QIAs, are stored as quantum states and chained together by entanglement \cite{cortese2018loading}. The QIAs are generated by issuers, and then verifiers perform QIA authentication according to user requirements. After, the users who want authorized actions from verifiers need to perform the authentication and proof process.

Attestations are evidence of statements about the user's identity, and they can be hash values, encrypted messages or bit strings, etc. QIAs in QBIF can be created by converting them into quantum bits \cite{wang2009theoretical}. A user can have multiple QIAs for different identity properties.
Before using the QIA, it should be double-signed by the owner and the issuer, which are stored in the chain of quantum states. The signature is done between them using the private key signature of both parties. The use of the issuer's private key signature illustrates that the issuer agrees with the user's attested statement, while the use of the user's private key signature indicates ownership of the attestation.

\subsubsection{Identity Management on Quantum blockchain} 
QBIF provides distributed identity management for coordinating quantum network nodes (including users, issuers, and verifiers). These nodes are interconnected through a quantum communication channel. Each type of quantum network node has a copy of a quantum state chain that holds signed QIAs in chronological order. Once a QIA is kept in the blockchain, the user sends the location of the signed QIA to initiate the verifier's operation. Then, the verifier validates the signed QIA using the public keys of the user and the issuer to determine the authenticity of the QIA. It's worth noting that the validity level of the validity can be defined by the issuer using the password, online, etc. In addition, the validity level can be integrated into the QIA by adding additional quantum bits. 

This quantum blockchain identity management deals with the primitive quantum blockchain structure, where some data transfer and payment transactions can take place off-chain. With the development of quantum technology, a hybrid network of classical and quantum communications might be the best way to transition from classical to quantum Internet. Meanwhile, the proposed framework faces some limitations and challenges, but it shows an obvious path to realize the decentralization of quantum blockchain.

\subsection{Scalability}

Scalability is a central issue of quantum blockchain that has attracted the attention of researchers. It aims to increase transaction throughput (i.e., transactions per second) while keeping the resources required for a party to participate in the consensus mechanism roughly constant and maintaining security against adversaries.

Traditional blockchains, such as Bitcoin and Ethereum, can currently only process 10 transactions per second. Quantum information is intrinsically well-suited to solve this problem. With the help of quantum technology, classic blockchains can leverage quantum cryptography to give the solution to the scalability problem. Moreover, smart contracts in blockchain can be combined with quantum tools, specifically quantum lightning, to design a decentralized payment system that addresses the scalability of transactions. Quantum lightning was proposed by Zhandry \cite{zhandry2021quantum} and inspired by the concept of collision-resistant quantum money \cite{lutomirski2011component}. The quantum lightning scheme for payment systems consists of a generation process to generate quantum banknotes and a verification process to validate the quantum banknotes and assign serial numbers to guarantee security. Here, the security guarantee means that no generation process can produce two banknotes with the same serial number unless the probability is negligible. This property prevents anyone from cloning banknotes. As such, it requires a mechanism to regulate the generation of new valid quantum banknotes.

Coladangelo \emph{et. al} proposed the smart contract is used to keep track of valid serial numbers \cite{coladangelo2020quantum}. Any party is allowed to deposit coins $d$ into a smart contract with specific parameters, and with an initial value of a serial number state variable. We can consider that the quantum banknote with the chosen serial number ``gets" the value of $d$. The integration between smart contracts and quantum lightning improves the speed of payment transactions, further tackling the problem of scalability.

Moreover, a consensus mechanism based on quantum cryptography, and quantum measurement is designed to reduce throughput and latency \cite{bhavin2021blockchain}. This scheme aims to determine the success of miners’ mining while avoiding consuming a large number of computing resources and mathematically complex calculations in the mining competition. Therefore, it reduces resource consumption. Additionally, quantum teleportation mainly consists of classical information transmission and quantum gates performing. It takes a very short time, while the transmission of quantum states is completed instantaneously. Therefore, the scalability of this scheme is high and the delay is small theoretically.

\subsection{Privacy of Storage}

Traditional blockchain systems store transaction data in the form of a distributed ledger, with each node storing copies of all data, which raises the storage problem. With the development of quantum technology, numerous researchers are looking for suitable quantum/post-quantum encryption and verification algorithms to improve standard distributed storage blockchain systems. 

\subsubsection{Post-quantum Based Distributed Storage} 

Existing research work proposes some secret sharing schemes to overcome the problem of distributed storage. Secret sharing, as one of the most important encryption protocols, is used to ensure the privacy of data \cite{blakley1979safeguarding,shamir1979share}.

A secure threshold-based verifiable multi-secret sharing scheme is proposed that does not require a private channel and shares two secrets simultaneously in a single sharing process among blockchain nodes based on post-quantum encryption algorithms \cite{mesnager2020threshold}. The scheme is then applied to a distributed storage blockchain system for the distribution and private storage of data blocks. Before distributing the data among the nodes of each block, we encrypt the data blocks with the AES-256 encryption algorithm. Meanwhile, this scheme shares secret keys and hashes of the blocks among the blockchain network nodes simultaneously. After, the encrypted data blocks are encoded by Reed-Solomon codes and shared among these blockchain nodes.

In addition, several secret sharing schemes based on lattices have been studied in \cite{amroudi2017verifiable,pilaram2015efficient}. Based on these post-quantum encryption schemes, the distributed storage blockchains also greatly improve the storage costs of traditional blockchains \cite{raman2018distributed,kim2018efficient}. The scheme also brings desirable features to distributed storage blockchain systems, such as (quantum-secure) authentication algorithms and secret communication without private channels. Notably, the flexible threshold parameters of the proposed scheme eliminate the shortcomings of the previously distributed storage blockchain in terms of recovery communication cost and robustness.

\subsubsection{Implementation Tutorial for Storage} 

Taking electronic medical records (EMRs) as an example, this section introduces how to realize distributed data storage with blockchain-based on quantum technology.

\begin{figure*}[!t]

\centering

\includegraphics[width=9cm]{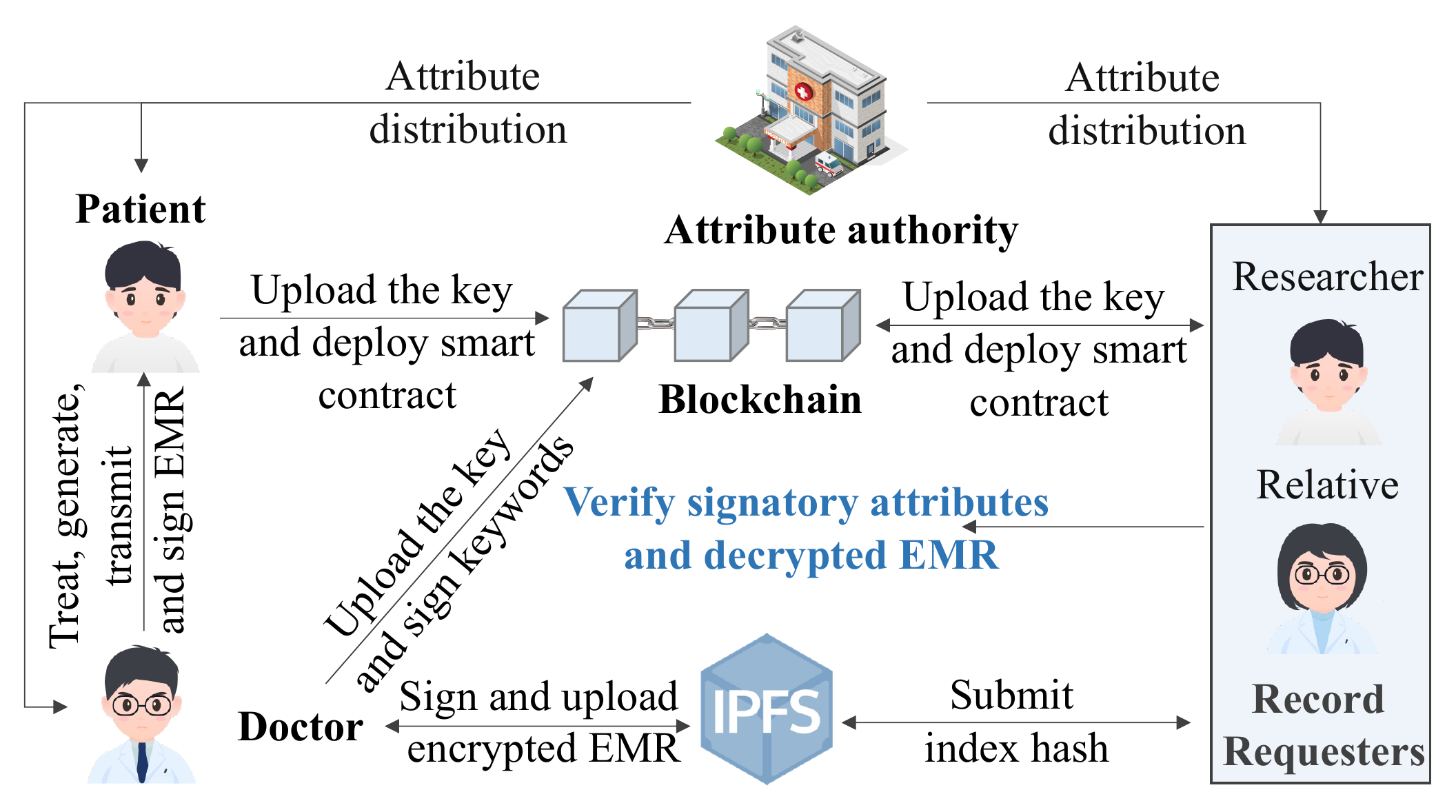}

\caption{The system architecture of AQ-ABS}

\label{fig:implement}

\end{figure*}

Along with the unceasing progress of medicine, it has become a popular trend to implement EMRs to improve the efficiency and reliability of medical services \cite{zeng2021efficient,he2012toward}. Nonetheless, EMRs are stored independently in hospitals and healthcare facilities, which leads to sharing problems. In addition, highly sensitive EMRs are vulnerable to being tampered with, posing a threat to privacy and security.
An anti-quantum attribute-based signature (AQ-ABS) is developed to address the above challenges \cite{chen2022aq}. AQ-ABS combines IPFS with the consortium blockchain to encrypt and sign EMRs, then store them in a secure and distributed file storage system, i.e. interPlanetary file system (IPFS). After, the index hashes generated by IPFS and keywords are re-signed and stored in the blockchain. As shown in Fig. \ref{fig:implement}, the system implementation process for EMRs shared plans are divided into the following phases.

\begin{itemize}
\item \emph{System Initialization:} This step includes registration and security parameter settings. These attributes and secret keys are uploaded to the blockchain.
\item \emph{EMR Generation:} When a user is treated for a disease at one hospital, it can consult the hospital and generate an EMR, as in Figure 1. If the user needs to be transferred to another hospital for treatment, the physicians at that hospital must be familiar with the patient's previous medical experience. That is, each physician is responsible for uploading the EMR for data sharing. In this case, the EMR record is generated during the doctor's treatment. 
\item \emph{Signature Generation and Uploading:} The physician responsible for the consultation sends an EMR to the patient to confirm the accuracy of the EMR. 
After receiving the EMR with the user's signature, the physician encrypts the EMR by \cite{brost2020threshold} and signs it to ensure unforgeability based on AQ-ABS. In addition, a combination of IPFS and consortium blockchain is used to achieve secure EMR sharing. A physician uploads an encrypted EHR with a double signature to IPFS to obtain an index hash. The physician then signs the index and keywords of the EHR and uploads them to the blockchain for distributed storage, while the patient deploys a smart contract to the blockchain for access control of the EHR.
\item \emph{Query and Signature Verification:} This step allows the user to download the EMR and verify the signature. As a patient's relevant researchers wish to access the electronic medical record, they first provide their attributes to the blockchain, which is verified by the smart contract, and search the index. If it is accessible, the blockchain returns the index value with the signature, otherwise, it rejects it. After receiving the signed index, the record requester verifies the signature, and if it passes the verification, submits the index to IPFS for EMR.

\end{itemize}

\emph{\textbf{Lessons learned:}}In the survey, we explore the external performance improvement in the quantum blockchain, including quantum blockchain economy, decentralization, scalability, and privacy of storage. This section explains that quantum information technology can promote the performance of blockchain, further improving the underlying infrastructure of Web 3.0, accelerating the construction of Web 3.0 application scenarios, and exploring the implementation of Web 3.0 application scenarios. Overall, we summarize the above-mentioned works of external performance improvement in Table \ref{tab:external}.
% \subsection{Security}

\section{ Applications and Implementation Tutorials of Quantum Blockchain in Web 3.0}
\label{Applications}

\begin{figure*}[!t]

\centering

\includegraphics[width=13cm]{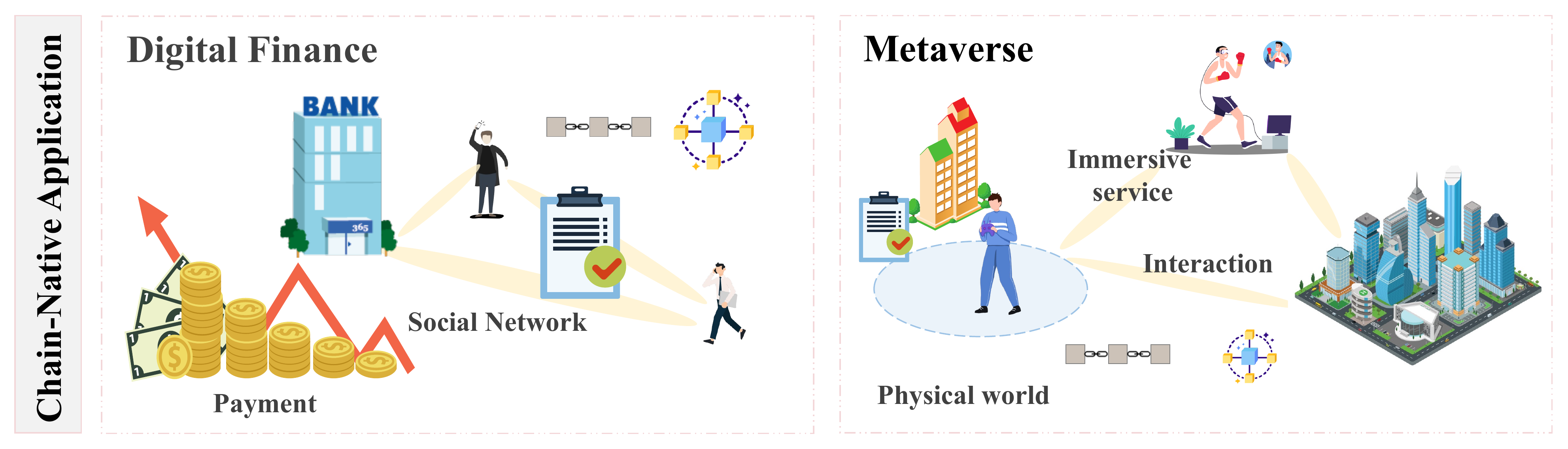}

\caption{  Chain-native services  of quantum blockchain in Web 3.0}
\label{fig:Chain_services}

\end{figure*}

\begin{figure*}[!t]

\centering

\includegraphics[width=13cm]{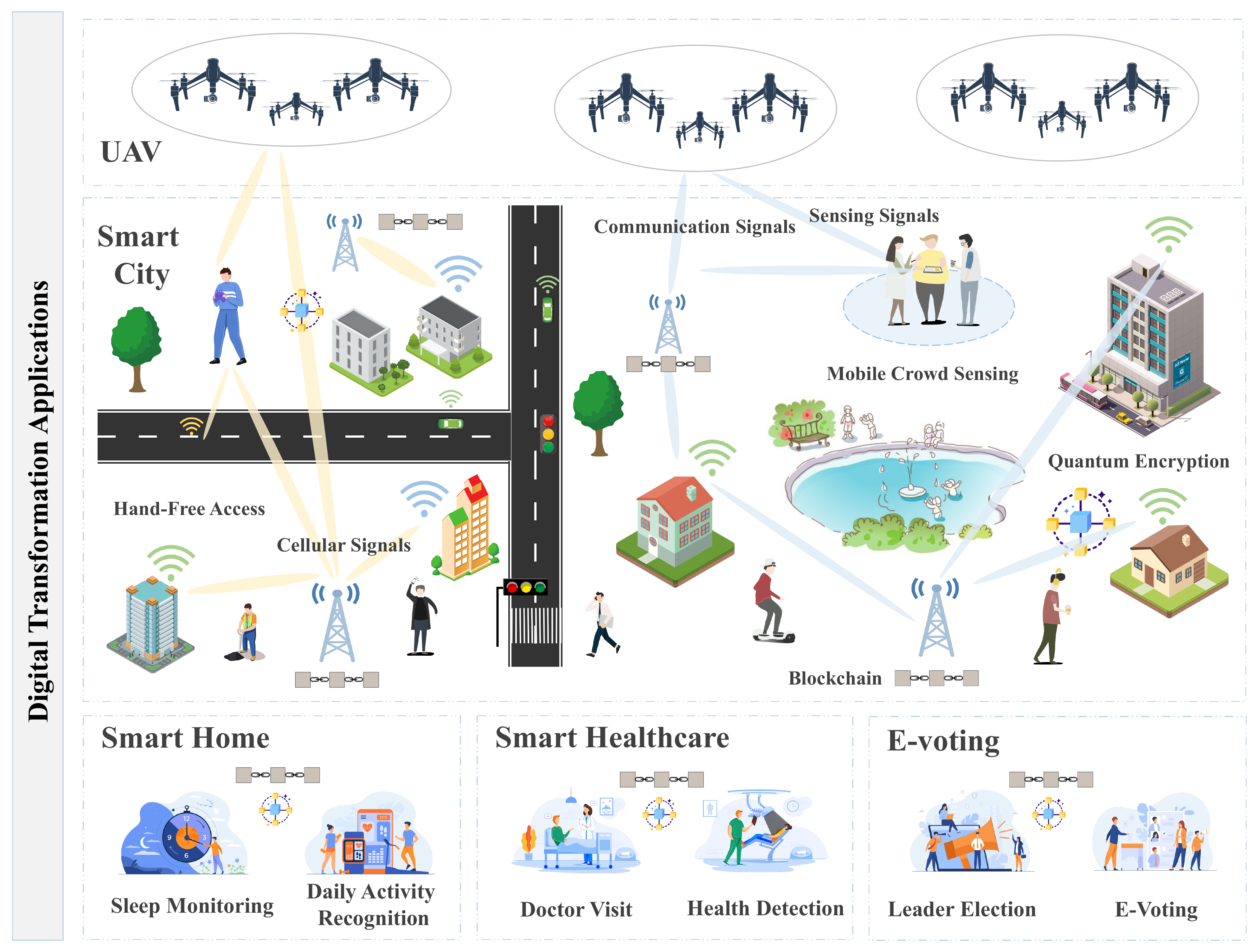}

\caption{ Digital transformation applications of quantum blockchain in Web 3.0}
\label{fig:Digital_Applications}

\end{figure*}

This section briefly introduces some applications and implementation tutorials of quantum blockchain in Web 3.0, These applications include chain-native services and digital transformation applications, as shown in Fig. \ref{fig:Chain_services} and Fig. \ref{fig:Digital_Applications}, respectively. 

\subsection{Chain-Native Services}
%\subsubsection{Digital Market}

\subsubsection{Digital Finance}

Modern digital finance is more concerned with features such as high reliability and security. As the major blockchain-based innovation, non-fungible token (NFT) is a non-fungible digital asset that cannot be replaced by its equivalent one \cite{Wang21NFT}. Further, it connects the physical world with the real world through numerous applications, such as identity, private equity transactions, intellectual property assets, digital collection, etc. To ensure the security of quantum NFT, the work in \cite{Pandey22NFT} proposes a novel protocol on a blockchain where the traditional ledger and hash functions are replaced by entanglement. 
This Quantum NFT protocol generally has three steps, including the creation of a block of NFT, the creation of token, and the verification of the blocks.

\subsubsection{Metaverse} 
Based on decentralized finance, Metaverse operates with its own economy, drawing escalating attention to the next-generation Internet. Generally, Metaverse is defined as a shared virtual space formulated by sensing, communication, and computing technologies, etc. Meanwhile, other technologies, such as blockchain, and quantum computing, are proving popular for a number of key Metaverse applications \cite{QuantumImpact}, as shown in the following.

\begin{itemize}

\item \emph{Security:} As more interactions between the physical and the virtual world in the Metaverse are captured, quantum-resistant blockchain technology is designed for all transactions and commerce. This also ensures that blockchain transactions remain safe against algorithms such as Shor’s algorithm \cite{Quantumplaymeta}.
\item \emph{Computation:} One of the significant advantages of quantum blockchains is that they are computationally efficient, ensuring the increased effectiveness of Metaverse applications. Concretely, quantum blockchain enhances the computation and overall experience, which makes the Metaverse user-friendly. 
\item \emph{Randomness:} To ensure that the blockchain-based Metaverse system is not manipulated by occupants and algorithms, a certain amount of quantum randomness is needed. Instead of pseudo-random numbers, a qubit series is leveraged to generate random bits, further preventing the metaverse from being utilized in unscrupulous ways. Currently, some companies, such as Quantum Dice \cite{QuantumDice}, are working in the field of quantum random number generation.

\end{itemize}

\subsection{Digital Transformation Applications}

\subsubsection{Smart Cities}

Smart cities aim to develop protocols and technologies to improve the quality of People's daily life. In this regard, quantum cryptography helps to eliminate the risk of data storage and transmission. The work in \cite{ABDELLATIF2021102549} designs a novel authentication and encryption protocol using quantum walks (QIQW). The QIQW protocol enables IoT nodes in smart cities to securely share data and have full control of their records. Meanwhile,  the novel blockchain framework based on QIQW is able to resist probable quantum attacks. 

In addition, post-quantum blockchain (PQB) is another solution to improve the security of transaction processing in the blockchain. To construct a lightweight PQB transaction in a smart city, the work in \cite{CHEN2021102780} embeds a post-quantum identity-based signature scheme and the InterPlanetary File System (IPFS). Some post-quantum solutions \cite{PostAzzaoui}, such as lattice-based cryptography, quantum key distribution, and entanglement, are experimentally easy to counter quantum attacks. This PQB architecture makes smart cities safer and provides a better place to live.

\subsubsection{Smart Healthcare} Smart healthcare aims to provide patient-centric healthcare services by secure data collection, efficient data processing, and systematic knowledge extraction \cite{Baker17Smart}. However, maintaining the security and privacy of stakeholders is a challenge for traditional healthcare systems. It would be helpful to develop a quantum blockchain, and quantum electronic medical records (QEMRs) using the security of quantum cryptography itself \cite{Chen2022ABS}. To defend against quantum attacks on traditional Healthcare systems, the quantum blind signature is leveraged for key distribution during the block creation \cite{BHAVIN2021102673}. And various smart contracts help users to access the medical data of the blockchain with their defined role. Replacing traditional encryption signature algorithms with quantum authentication systems, a quantum electronic medical record protocol is designed \cite{QU2022942}. This protocol tracks every medical record while guaranteeing the security and privacy of EMRs in medical systems. Additionally, quantum computing is able to support large operations and complex computation on uninstructed data, further improving the quality of service provided by intelligent healthcare systems \cite{ELAZZAOUI2022103304,Amin2022}. 

\subsubsection{E-voting }

Electronic voting, i.e., E-voting, is a voting process in which votes are cast and counted with computer assistance. Recently, the possible integration of blockchain and quantum technology could facilitate the implementation of E-voting with much greater transparency and security. Sun \emph{et. al} propose a voting protocol based on the quantum blockchain \cite{sun2019simple}. This protocol not only simplifies the task of E-voting but also satisfies many significant properties, such as anonymity, verifiability, eligibility, fairness, etc. However, this protocol is limited in that it cannot be applied beyond a certain number of users. In addition, denial-of-service attacks can easily be used by one voter against another. To circumvent the possible attacks by upcoming noisy intermediate scale quantum (NISQ) computers. 
By enhancing the advantages offered by blockchain with the quantum protocols \cite{Mishra2022voting}, such as quantum random number generators, quantum key distribution, etc., an anonymous voting scheme is developed to circumvent the possible attacks from the upcoming Noise Intermediate Scale Quantum (NISQ) computers. The quantum protocols enable the E-voting systems to secure against any adversary, whereas there exists the central authority in the elections, which may adversely affect the voting outcome.
Hence, it is necessary to consider the audit function in the E-voting protocol in order to ensure the efficiency and fairness of the voting process \cite{Gao2019Quantum,Chow2007Ring}. 

\subsubsection{Smart Home and UAV} As prominent representatives in the Internet of things (IoT), smart homes, and UAVs have surged forward in the past decade, improving our daily lives. 

On the one hand, smart homes aim to offer intelligent sensing and collaborative control in the home. It is noted that security is an important problem before it can be widely adopted. One plausible solution to security breaches is to introduce blockchain and quantum technology to increase safe communication among smart devices \cite{Post21Chen}. To this end, a consortium blockchain can be formed to provide smart home services through smart contracts. Meanwhile, the post-quantum signature schemes, such as pqNTRUsign, are designed to replace existing ECDSA signature schemes, helping to protect the blockchain-based smart home services from a potential quantum computer attack. 

Unmanned aerial vehicles (UAVs), on the other hand, are seen as a potential asset due to their broader application in scope \cite{Ralegankar22UAV}. A number of communication protocols are needed to support the successful flight of UAVs. In UAV communications, blockchain-based quantum systems could be developed to solve security issues in the distribution of critical information. Thereinto, a quantum cryptography-based architecture leverages the properties of quantum cryptography and beyond 5G networks to enable drone communication more shielded from data transmission and to achieve high security and efficient data transmission. 

\subsection{Implementation Tutorials of Quantum Blockchain in Web 3.0}
In this section, we provide a tutorial on the implementation of Quantum Blockchain in Web 3.0, which consists of the following four phases.

\subsubsection{Generation of Quantum Entropy and Post-Quantum Certificates }

In cryptographic processes in quantum blockchain, sufficient randomness prevents a large number of malicious attacks. The generation of quantum randomness takes advantage of the non-deterministic property of quantum mechanics \cite{zheng2020bias,herrero2017quantum}. Considering the distributed nature of the blockchain, ideally, each blockchain node should have its own local source of quantum entropy. Then, quantum communication protocols need to be designed so that blockchain nodes create a quantum secure tunnel between themselves and the entropy distribution point to ensure secure communication. In this respect, the entropy source creates the first key, splits it into parts, and delivers it to the blockchain node through various TLS channels. 

Once a blockchain node accesses quantum entropy on demand, the entropy is consumed by OpenSSL, the cryptographic framework for applications that use TLS/SSL. After that, the blockchain nodes encapsulate the communication with other nodes using the post-quantum key generated by the signature algorithm and sign the transactions broadcasted to the blockchain \cite{alagic2020status}.

\subsubsection{Nodes Communication based on Quantum Cryptography}

Communication between nodes is established through quantum protocols in blockchain, making nodes' communication quantum-resistant. It can be achieved by adding peer-to-peer TLS tunnels that have been modified to support post-quantum keys. Since this is built on a TLS connection that is insensitive to key length, other post-quantum algorithms can be used, unlike blockchain transactions \cite{FalconGitHub}.

\subsubsection{ Signature of Transactions}

The implementation of this phase requires us to pay attention to each specific blockchain network. Some blockchain protocols recognize different encryption algorithms or have the flexibility to incorporate new ones. In this case, we need to add quantum signatures to transactions broadcast to the network without modifying the blockchain protocol by developing relay signers and meta-transaction signature models \cite{EIP-155}.

\subsubsection{ On-chain Verification}
When a blockchain writer node adds a post-quantum signature to a meta-transaction and broadcasts it to the network, it needs to be verified on-chain. Once a relay hub smart contract is verified as a transaction target, each node extracts the original transaction information to verify the signature. The process involves three steps:

\begin{itemize}
\item  The blockchain node receiving the meta-transaction checks the sender. Next, they verify that the public key derived from the meta-transaction signature controls the decentralized identifier (DID) of the node that generated the transactions \cite{W3C}.
\item If the previous step completes successfully, the node invokes the DID registry again, after which it resolves the post-quantum public key associated with the DID and the blockchain public key verified in the previous step.
\item After, the post-quantum public key, post-quantum signature, and original transaction are obtained from the DID registry. Further, each node verifies the post-quantum algorithm on-chain.
\end{itemize}

With the above steps, blockchain nodes add meta-transactions to the transaction pool and copy them to other nodes so that the validator receives them and adds them to the next block.

%To react to the risk posed by forthcoming quantum computers, a number of quantum voting protocols have been

\emph{\textbf{Lessons learned:}} Quantum blockchain applications improve the security, reliability, and efficiency of blockchain-based solutions in various areas such as chain-native services and digital transformation applications. However, the implementation of quantum blockchain applications is still in its early stages and there are many open issues to be solved before quantum blockchain applications can be widely deployed.

\section{Future Research Directions and Open Issues}
\label{Challenge}

Although blockchain holds great promise in Web 3.0 with the help of quantum information technology, there are ongoing challenges in future research. In this section, we discuss several open issues and explore future directions, including quantum crypto-economics, quantum intelligence, and transaction intelligence.

\subsection{Migration to Quantum and Post-quantum Blockchain}
Since classical cryptography is expected to be broken by quantum computers in the near future, the migration to quantum and post-quantum cryptography-based cryptosystems needs to be efficient and smooth. In this process, the legacy blockchains of Web 3.0 also need to be migrated to quantum and post-quantum blockchains. However, this migration is not just about adding QKD and post-quantum cryptographic algorithms to existing blockchains. First, all existing encrypted data based on traditional cryptographic algorithms, such as RSA, can be deciphered by quantum computing, which means that the data in traditional open systems might need to be re-encrypted. Since the existing data can be intercepted in transit for future cracking. Finally, there is the issue of prioritizing network technology facilities during the transition process. Due to limited migration resources, the most critical Web 3.0 services need to be migrated to quantum and post-quantum blockchains at the earliest. However, how to decide the priority in the migration to quantum and post-quantum blockchains remains to be studied.

\subsection{Internal Technologies and External Performance}
% Research challenges
\subsubsection{Standardization for Quantum and Post-Quantum Blockchain}

Standardization of quantum and post-quantum blockchains relies on the standardization of quantum networks and post-quantum cryptography. After the standardization Web 3.0 can be put into industrial production on a large scale to rapidly increase the scalability and service diversity of digital ecosystems. First, the standardization of quantum networks can lead to efficient and absolutely secure quantum cryptography schemes for quantum blockchains. Then, the results of the standardization of post-quantum blockchains might be determined by a post-quantum cryptography competition conducted by NIST. In this standardization competition, NIST conducts extensive testing, evaluation, and comparison of the advantages and disadvantages of different PQC schemes in terms of efficiency, security, compatibility, and scalability. As of now, the standardization of post-quantum cryptography can be completed within a few years, while the deployment of quantum networks might take decades to develop.

\subsubsection{Quantum Intelligent Blockchain}

The quantum intelligence blockchain uses quantum machine learning to improve the internal technology and external performance of the quantum blockchain. For example, quantum machine learning can be used to design the incentive mechanism of the quantum blockchain, which can be trained to obtain the optimal incentive. In addition, quantum machine learning can also be used to design quantum circuit structures for running quantum consensus algorithms. Finally, quantum machine learning can be used to learn routing algorithms and block propagation algorithms in quantum networks to improve the throughput of quantum blockchains while maintaining a decentralized state. Overall, quantum intelligent blockchain is promising in improving the efficiency and scalability of the quantum and post-quantum blockchain in Web 3.0.

\subsubsection{Quantum Decentralized Digital Identity}

Decentralized digital identities allow users to manage their digital assets autonomously, securely, and transparently in Web 3.0. In quantum and post-quantum blockchains, users can obtain verifiable, privacy-protected, and revocable digital identities without relying on third-party identity providers. Providing digital identities through QKD-based cryptography provides unconditionally secure data identity authentication and transmission. However, users need access to the quantum network to use applications in Web 3.0 through quantum digital identities. Therefore, the future research direction for quantum digital identity needs to be based on the improvement of quantum network throughput and coverage, so that users can protect their privacy and ownership through quantum communication in Web 3.0 according to different scenarios and needs.

% Decentralized Digital Identity 是用户在 Web 3.0中的

\subsubsection{Quantum Decentralized Finance}

By building Web 3.0 decentralized exchanges on quantum and post-quantum blockchains, traders can complete the transmission of transaction information under an absolutely secure communication link to prevent man-in-the-middle attacks. Quantum networks can be deployed between major exchanges to reduce the latency of information exchange between exchanges. In this way, the information gap between exchanges is dramatically reduced, thus avoiding some arbitrage activities due to network dominance. In addition, Web 3.0 allows its participants to borrow and lend without collateral, secured by quantum cryptography and post-quantum cryptography. Therefore, future research directions could focus on how to design highly liquid lending protocols in Web 3.0 through quantum and post-quantum cryptography to enhance the sustainability of the decentralized ecosystem.

% \subsection{External Performance}
% Future directions

\subsection{Convergence of Classic-Quantum Networking and Computing}

In Web 3.0, blockchain-based computation-power networks connect heterogeneous computational resources in existing classic and quantum networks to support decentralized applications and services. Based on quantum blockchain, quantum computing and quantum networks can be integrated into the core of Web 3.0. Therefore, Web 3.0 can allocate these computing and networking resources to facilitate internal technologies and improve external performance. For example, Web 3.0 with the convergence of classic-quantum networking and computing can leverage quantum computing to handle more complex and large-scale optimization, simulation, and machine learning problems. Furthermore, it can provide faster, more stable, and more scalable network connectivity for decentralized applications through quantum networks. While combining quantum and classical computing-power networks can bring more potential to Web 3.0 in driving digital transformation, further research is needed to design appropriate protocols and algorithms to divide, schedule, synchronize, and converge quantum-classical computing and networking.

\emph{\textbf{Lessons learned:}} Quantum blockchain is still in the early stage of development, and it needs continuous exploration and innovation in terms of reliability, security, efficiency, and scalability. In this section, we give some future directions to help build the resilient Web 3.0 infrastructure.

\section{Conclusion and Discussions}
\label{Conclusions}

Given the complementary advantages between quantum information technologies and blockchain, it is clear that the integrated quantum blockchain shows a way to implement distributed Web 3.0 infrastructures. In this article, we have addressed the quantum blockchain and thoroughly examine most of the developments in Web 3.0. Specifically, we have described the integration mainly in terms of two aspects, namely, optimizing internal technologies and improving external performance. Then, we presented some potential quantum blockchain applications and tutorials for implementation in Web 3.0. Finally, We have discussed some important challenges and research directions that open new horizons for Web 3.0 development.

So far, quantum blockchain is still in its infancy and is expected to play a growing role in meeting the diverse needs of Web 3.0. This work provides a comprehensive, systematic, and in-depth introduction to the quantum blockchain for its further research and applications in Web 3.0.

\bibliographystyle{IEEEtran}
\bibliography{sample-base}


\end{document}